\documentclass[twocolumn]{autart} 

\usepackage{tikz}
\usepackage{tikz-network}
\usepackage{amsmath}
\usepackage{amssymb}
\usepackage{amsfonts}
\usepackage{graphicx}
\usepackage{enumitem}
\usepackage{mathrsfs}
\usepackage{float}
\usepackage{cite}
\usepackage{tikz}
\usetikzlibrary{matrix,arrows,decorations.pathmorphing}
\usepackage{verbatim}
\usepackage{mathtools}
\usepackage{caption}
\usepackage{subcaption}
\usetikzlibrary{arrows}
\usepackage{tabularx}
\usepackage{xcolor}
\usepackage{inputenc}
\usepackage[T1]{fontenc}

\newtheorem{lemma}{Lemma}
\newtheorem{remark}{Remark}
\newtheorem{assumption}{Assumption}

\newtheorem{assumptionalt}{Assumption}[assumption]

\newenvironment{assumptionp}[1]{
  \renewcommand\theassumptionalt{#1}
  \assumptionalt
}{\endassumptionalt}

\DeclareMathOperator*{\Dg}{\mathsf{diag}}

\renewenvironment{pf}{{ {\textbf{Proof}}$\;$}}{\par}

\begin{document}
\begin{frontmatter}

\title{Discrete-time Layered-network Epidemics Model with Time-varying Transition Rates and Multiple Resources} 

\author[rugjbi]{Shaoxuan Cui}\ead{s.cui@rug.nl},    % Add the 
\author[fzl]{Fangzhou Liu}\ead{fangzhouliu.ac@gmail.com},               % e-mail address 
\author[rugjbi]{Hildeberto Jard{\'o}n-Kojakhmetov}\ead{h.jardon.kojakhmetov@rug.nl},  % (ead) as shown
\author[rugenteg]{Ming Cao}\ead{m.cao@rug.nl} 

\address[rugjbi]{Bernoulli Institute, Faculty of Science and Engineering, University of Groningen, The Netherlands}  % Please supply                                              
\address[fzl]{Research Institute of Intelligent Control and Systems, School of Astronautics, Harbin Institute of Technology, China}             % full addresses
\address[rugenteg]{ENTEG, Faculty of Science and Engineering, University of Groningen, The Netherlands}        % here.

\journal{}
          
                             % keyword list or with the 
                                          % help of the Automatica 
                                          % keyword wizard

\begin{abstract}                          % Abstract of not more than 200 words.
This paper studies a discrete-time time-varying multi-layer networked SIWS (susceptible-infected-water-susceptible) model with multiple resources under both single-virus and competing multi-virus settings. Besides the human-to-human interaction, we also consider that the disease can diffuse on different types of medium. We use \emph{resources} to refer to any media through which the pathogen of a virus can spread, and do not restrict the resource only to be water. In the single-virus case, we give a full analysis of the system's behaviour related to its healthy state and endemic equilibrium. In the multi-virus case, we show analytically that different equilibria appear driven by the competition among all viruses. We also show that some analytical results of the time-invariant system can be expanded into time-varying cases. Finally, we illustrate the results through some simulations.
\end{abstract}

\end{frontmatter}

\section{Introduction}
During the special time of Covid-19 crisis, the study of epidemic models has drawn a lot of attention from communities of applied mathematics \cite{bayraktar2021macroeconomic}, physics \cite{schlickeiser2021analytical} and biology \cite{djordjevic2021systems}, among others. Besides the spread of a disease, epidemic models can also be used to describe the  spreading process in a more general context, such as information diffusion in social networks~\cite{li2017survey} and computer virus dissemination \cite{lloyd2001viruses}, which lies in the forefront of information science to social science. Various models have been developed to characterize such a process, e.g., susceptible-infected (SI), susceptible-infected-susceptible (SIS), and susceptible-infected-recovered (SIR). An overview of these models is provided in \cite{mei2017dynamics,zino2021analysis}.

The SIS model originates form the early study of \cite{kermack1932contributions}. The SIS model is a classic epidemics model, in which an individual is either in the susceptible (S) state or infected (I) state. On one hand, a healthy agent becomes infected at some infection rate $\beta$ depending on the interactions with its neighbours. On the other hand, an infected individual can recover and return to a susceptible state at some healing rate $\delta$. In the real world, the disease spreading is a heterogeneous process in the sense that the way how individuals interact with each other may vary from group to group. To deepen the understanding of the spreading process among a large scale population, considerable efforts have been put on the research of networked SIS models, both in continuous time \cite{van2008virus,van2013homogeneous,khanafer2014stability} and discrete time \cite{pare2018analysis,prasse2019viral,liu2020stability}. These networked SIS models show great potential to describe the real propagation phenomena in biological processes and in innovation dissemination. For example, the discrete-time model \cite{pare2018analysis} has been validated by real
data in the area of virus spreading and innovation
diffusion. 

However, the aforementioned networked SIS model have several obvious disadvantages. One of the major drawbacks is that all models studied in \cite{van2008virus,van2013homogeneous,khanafer2014stability,pare2018analysis,prasse2019viral,liu2020stability} only consider time invariant-spreading parameters. This does not reflect the fact that most real spreading processes are time-varying. In \cite{pare2017epidemic}, a time-varying version of the SIS model is developed and analyzed to this end. {The discrete-time version is studied in \cite{9170816} with the strong assumption of the parameters changing periodically.} Furthermore, the models in \cite{van2008virus,van2013homogeneous,khanafer2014stability,pare2018analysis,prasse2019viral,liu2020stability,pare2017epidemic} are restricted to the single-virus setting, while ignoring the fact that multiple viruses can compete for spreading in real-life scenarios. Examples of the aforementioned framework include competing viruses in the biological process \cite{nowak1991evolution} or competing innovation products in  marketing \cite{prakash2012winner}. Along this line of research, a bi-virus continuous-time SIS model is proposed in \cite{liu2019analysis}, and it is shown that the competition among two viruses finally leads to either convergence to the healthy state, ``winner takes all'', or to coexistence of multiple viruses, depending on the condition. Its discrete-time version is proposed in \cite{pare2020analysis} and validated with real data. However, analysis of the endemic behaviours is not carried out. Combining the ideas of multi-virus and time-varying topology, the continuous-time multi-virus time-varying model is proposed in \cite{pare2021multi2} and the system's behaviour is further studied under the influence of  mutations and human awareness.

Note that the conventional networked SIS model only focuses on the direct human-to-human interaction. The indirect spreading of pathogens through other media (resources) is not taken into consideration. In this work, resource refers to media, and both words are used interchangeably. For instance, a waterborne disease can spread through the water supply and consequently infect the people who use such water resource \cite{kough2015modelling}. The disease can also diffuse through insects \cite{coutinho2010transgenesis} or over the ventilation system \cite{mihai2021overview}. Based on the SIS type disease spreading, by adding one water counterpart $W$ (not restricted to be water resource and can be other resource) into the conventional SIS model, the continuous-time SIWS model\footnote{In this context, SIWS stands for susceptible-infected-water-susceptible, and the use of $W$ should be distinguished from waning immunity, which has also been considered \cite{chaves2007loss,lavine2011natural,jardon2021geometric}.} is proposed in \cite{liu2019networked}. However, in this work, only one single water resource and one single virus is considered in the model. Recently, the model has been further expanded into a multi-virus single-resource SIWS model in \cite{janson2020networked}, and a single-virus SIWS model coupled with an infrastructure network (multiple resources involved) in \cite{pare2021multi}. By considering a time-varying topology, the single-virus and single-resource time-varying SIWS model is studied in \cite{gracy2021analysis}.

Although it is evident that an intensive effort has been put on the study of the continuous-time SIWS model, to the best of our knowledge, no previous study is dedicated to the modeling and analysis of a discrete-time SIWS model. It is very important to fill this gap, since real data is frequently sampled in a discrete manner, and therefore a discrete-time model could be more suitable for applications. {Furthermore, as indicated by the survey paper \cite{pare2020modeling}, the majority of parameter estimation algorithms are designed only for discrete-time models. Therefore, to deal with the real data, in several works such as \cite{pare2018analysis,pare2020analysis}, a discrete-time model has been used to approximate the continuous-time model. Moreover, it is also shown that the discrete-time models usually work well with real-data \cite{pare2018analysis,pare2020analysis}. However, there is no general guarantee that the discrete-time model will have the same properties as the continuous-time model, indicated by \cite{prasse2019viral}. That is why one needs to develop an appropriate discrete-time epidemic model to capture the qualitative behaviour of a viral spread.}

Furthermore, no previous work deals with the variation of multi-virus, multi-resource and time-varying spreading parameters at the same time. In real spreading processes, transition rates are time-varying and multiple viruses diffuse via different kinds of media. As an example, just consider the most recent scenario of Covid-19, where the Omicron and Delta variants compete for spreading \cite{trigger2022two}. The spreading parameters of Covid-19 is no doubt time-varying and spatially-heterogeneous. Besides, Covid-19 pathogen is confirmed to be found in wastewater \cite{farkas2020wastewater} and different types of surfaces such as trees, buildings, and vehicles \cite{kumar2020artificial}. {In the medical or environmental sciences, many works \cite{hillary2020wastewater,daughton2020wastewater,bogler2020rethinking,robotto2022wastewater} emphasize the importance of monitoring Covid
infection levels by wastewater surveillance. However, model-based analytical tools for this kind of data is still missing. The work \cite{hillary2020wastewater} indicates that subsequent risk analysis and modelling regarding the wastewater surveillance is more than important to understand the dynamics of viral outbreak.} These aforementioned reasons motivate us to develop a multi-virus, multi-resource  discrete-time SIWS model with time-varying spreading parameters at the same time to have a more general setting to better characterize the real spreading process.

%Last but not least, how to control the pandemic, namely, in epidemiology, how to cut down the reproduction number based on mathematical model is no doubt a very hot topic in the control community. The general approach is to either remove some links in the graph or to boost the healing rate. The article \cite{zino2021analysis} offers an overview for these control strategies. Other than these two methods, it is shown by \cite{liu2019analysis, ye2020distributed} that a large series of distributed feedback control law can never stabilize the healthy state, however indeed mitigate the spreading to some extent. However, we will modify this control law to regulate our proposed system to converge to the desired endemic equilibrium in the context of virus marketing or information diffusion. Different from epidemiology, we aim at achieving endemic equilibrium with as high infection level as possible.

The contributions of the present paper are multiple: 1) We propose a discrete-time time-varying single-virus networked layered SIWS model with time-varying parameters and multiple resources. To the best of our knowledge, it is the first model to describe an SIWS type spreading process in discrete time. This single-virus model will serve as a foundation which allow us to deal with the modeling and analysis of our more general multi-virus model. 2) We further generalize the discrete-time  model into a setting of multiple competing viruses. Thus, it is also the first model to study multiple viruses, multiple resources, and time-varying parameters simultaneously in the SIWS type spreading process, which makes our model have a more general framework and potentially more suitable for applications. 3) We provide a full-system analysis for our proposed models with regard to both healthy state behaviour and endemic behaviour. {We improve results concerning the healthy state in the time-varying case. We address two open questions regarding the endemic behaviour, namely, 1) we assess the global stability of the dominant endemic equilibrium in the multi-virus discrete-time epidemics system, and 2) we prove the existence of endemic equilibrium under some condition in the time-varying case and }
{especially, }we study the question under which circumstances the time varying model converges to the endemic equilibrium. As mentioned above, all related works on time-varying SIS or SIWS model \cite{pare2017epidemic,pare2021multi2,gracy2021analysis} and multi-virus discrete-time SIS model \cite{pare2020analysis} do not address such {two questions}. %3) We review a series of preexisting control strategies designed for epidemics model and adapt them to our proposed model. 
4) We also present detailed numerical simulations highlighting our main results, and allowing us to discuss on more complex scenarios.

The paper is organized as follows: firstly, we  introduce some preliminaries and propose a single-virus discrete-time SIWS model in section \ref{sec::pre}. Then, we show our main results for the single-virus case in section \ref{sec:mrs}. Later on, in section \ref{sec:multi}, we extend and generalize the system into the multi-virus case and provide some analytical results. Afterwards, some numerical results are given to illustrate the main results in section \ref{sec:sim}. Finally, we make some discussion and draw relevant conclusions in section \ref{sec:con}.

\emph{Notations:} Let $\mathbb{R}$ and $\mathbb{N}$ be the set of real numbers and nonnegative integers, respectively. Given a matrix $M \in \mathbb{R}^{n \times n}$, $\rho(M)$ is the spectral radius of $M$, which is the largest absolute value of the eigenvalues of $M$. By $s_1(M)$ we denote the the largest real part among the eigenvalues of
$M$. For a matrix $M \in \mathbb{R}^{n \times r}$ and a vector $a \in \mathbb{R}^n$, $M_{ij}$ and $a_{i}$ denote the element in the $i$th row and $j$th column and the $i$th entry, respectively. For any two vectors $a, b \in \mathbb{R}^n$, $a \gg (\ll) b$ represents that $a_i >(<) b_i$, for all $i=1,\ldots,n$; $a > (<) b$ means that $a_i \geq (\leq) b_i$, for all $i=1,\ldots,n$ and $a \neq b$; and $a \geq (\leq) b$ means that $a_i \geq (\leq) b_i$, for all $i=1,\ldots,n$ or $a = b$. These component-wise comparisons are also applicable for 
matrices with the same dimension. The vector $\mathbf{1}$ ($\mathbf{0}$) represents the column vector or matrix of all ones (zeros) with appropriate dimension. The matrix $I$ represents the identity matrix with appropriate dimension. The notation $[n]$ denotes the set $\{ 1, 2, \dots, n \}$.

\section{Preliminaries and Single-virus Model Description} \label{sec::pre}
In this section, the background knowledge of graph theory are introduced. Then, we propose our discrete-time single-virus layered networked SIWS model. Finally, we give necessary assumptions for the model. This single-virus model will serve a starting point for our more general multi-virus model. 

\subsection{Preliminaries}
Consider a weighted directed graph $\mathcal{G}=(\mathcal{V},\mathcal{E},A)$ where 
$\mathcal{V}=\left\{ 1,2,\ldots,N\right\}$, and $\mathcal{E} \subseteq \mathcal{V} \times \mathcal{V}$ are the set of vertices and 
the set of edges, respectively. The matrix $A = [A_{ij}] \in \mathbb{R}^{N \times N}$ is the nonnegative adjacency matrix. The entry $A_{ij} > 0$ if and only if there exists an edge from node $j$ to node $i$. Otherwise, the entry $A_{ij} = 0$. The graph $\mathcal{G}$ is strongly connected, if between every pair of nodes there exits a path (a sequence of edges) between them. This strong connectivity reflects the fact that, ultimately, humans are interconnected with each other due to, for example, globalization. We recall that a square matrix is irreducible if it is not similar via a permutation to a block upper triangular matrix, and that the adjacency matrix $A$ of a graph is irreducible if and only if the associated graph $\mathcal{G}$ is strongly connected. By definition, a matrix $M$ is nonnegative if all the entries of $M$ are nonnegative. In the appendix \ref{app:prononneg}, some useful properties of nonnegative matrices are recalled.

%\begin{lemma}[\cite{berman1994nonnegative}, Corollary 1.5]\label{lemma:comp11}
%Let $A$, $B$ be nonnegative square matrices.
%If $0\leq A \leq B$, then $\rho(A) \leq \rho(B)$.
%If $0\leq A \leq B$, $A\neq B$ and $A+B$ is irreducible, then $\rho(A) < \rho(B)$.
%\end{lemma}

We now proceed to the description of our single-virus model.

\subsection{Model Description}

As background, let us first consider the conventional discrete-time SIS model proposed in \cite{pare2018analysis} based on a $n$-node human contact network $\mathcal{G}(A)$:
\begin{equation} \label{eq::sys_sis_dt}
 x_i(t+1) = x_i(t) + h \left( (1-x_i(t)) \sum_{j=1}^{n} \beta_{ij} x_j(t) -\delta_i x_i(t) \right),
\end{equation} 
where $x_i(t)$ denotes the infection probability of agent $i$ or the infection fraction of group $i$ at time step $t$; $\beta_{ij} := \beta_{i} A_{ij}$; $A_{ij}$ is the entry of the adjacency matrix of the corresponding $n$-node network $\mathcal{G}(A)$; $\beta_{i}$ and $\delta_{i}$ are the infection and healing rates of node $i$, respectively; and $h$ represents the sampling time.

{The introduction of a network structure and the adjacency matrix to epidemics models can be traced back to \cite{chakrabarti2008epidemic}. The weights of the  adjacency matrix $A_{ij}$ denote how close the pair-wise contact between groups is. It can be identified in at least two ways: firstly, traffic data can be directly employed \cite{parino2021modelling}; secondly, real epidemic data can be used to recover the graph connection \cite{prasse2020network,liu2021distributed}.}

%\hjk{In this context the fact that $a_{ij}\in\mathbb R_{\geq0}$ looses relevance no? wouldn't it suffice that $a_{ij}\in\left\{ 0,1\right\}$? If not, we need to explain why.}\csx{$a_{ij}\in\mathbb R_{\geq0}$ holds here. in the section of assumptions, we have some assumptions regarding $\beta_{ij} = \beta_{i} a_{ij}$, which also restrict $a_{ij}$. Maybe i should make a remark on it. however, it is not so important, because throughout the remaining paper, we will use $\beta_{ij}$ instead of $ \beta_{i} a_{ij}$}

Now, consider an SIS type disease which spreads over a human contact network consisting of $n$ groups of individuals %\hjk{do you mean a group of $n$ individuals?} \csx{no, it is just n groups of people.} \hjk{ok, I see that $x_i$ can stand for both probability of infection of an agent or the infection fraction of a group, which is weird, but ok}\csx{it is not wierd. if you treat the graph as a network of a country, each node is a population node of a city. ok, i will try to make it clear somewhere}
(agents) and its pathogen diffuses over $m$ different types of resources that are utilized by some agents in the human contact network. An individual can be infected with the disease either by its infected neighbour in the human contact network, or by the polluted resource the individual utilizes. In turn, the resource can be contaminated by the infected agents uses the resource. 
%Figure \ref{fig:net2} shows the state transition of an individual in the human contact network and the interplay between the individual and the resource it utilizes.

\begin{figure} 
    \centering
\begin{tikzpicture}
    \node at (0,0){
    \includegraphics[scale=0.75]{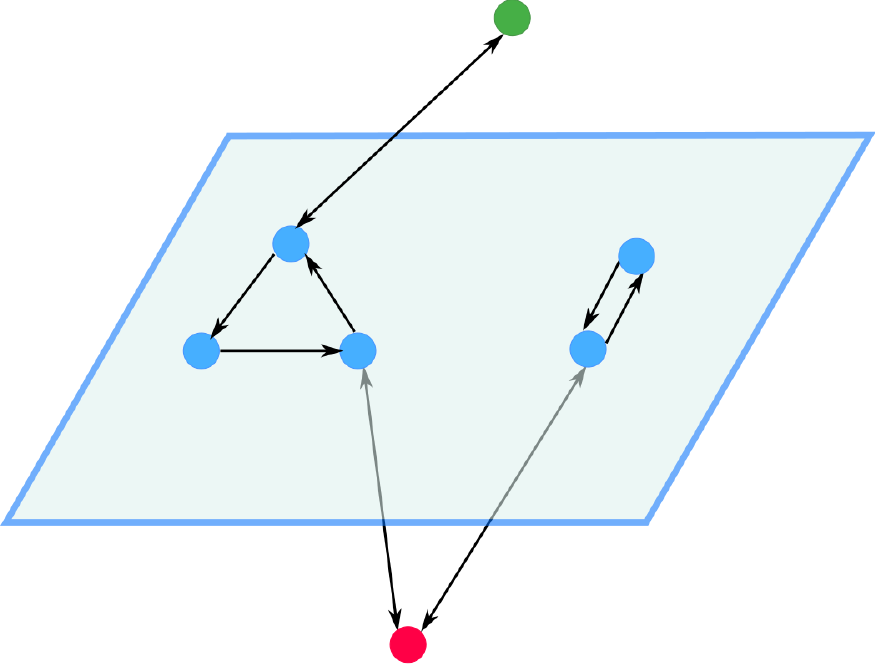}
    };
    \node at (2.2,2.4){\small{ventilation system}};
    \node at (-.8,-2.3){\small{water}};
    \node[blue!70!green] at (1.3,1.25){\small{human-contact network}};
\end{tikzpicture}
    \caption{An example of three layer spreading network. The human contact network is not assumed to be strongly connected. The resources are interconnected to at least one node in the human contact network. We assume that the whole equivalent graph when combining human contact network and multiple resources from different layer is strongly connected. This assumption is possible when the human contact network is not strongly connected, because the resource can serve as a medium connecting different groups of population.  The node in the human contact network  is the population node. It can denote a single individual or a group of people living in the same neighbourhood, city or province.}
    \label{fig:net1} 
\end{figure}
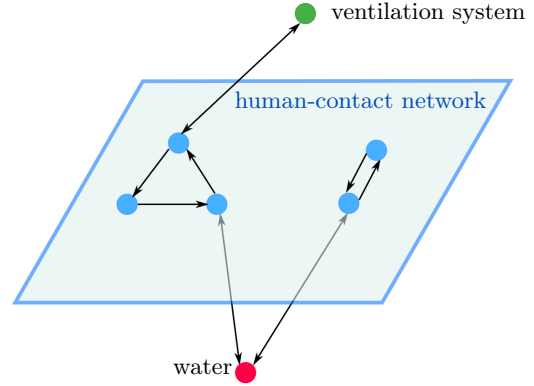

For the graph structure, the human contact network $\mathcal{G}(A)$ remains unchanged. We further introduce diverse types of resources (water, ventilation system, traffic system) and construct a multi-layer structure. Figure \ref{fig:net1} shows an example of such a graph structure.

By adopting the same pathogen dynamic in the resource and parameter modification introduced in \cite{liu2019networked}, we can get the following discrete-time SIWS model with multiple types of resources:
{\small\begin{align}
    x_i(t+1)&= (1-h\delta_i) x_i(t) \notag \\
    &+h\left[ (1-x_i(t)) \left(\sum_{j=1}^{n} \beta_{ij} x_j(t)
 +\sum_{j=1}^{m} \beta_{ij}^{w} w_j(t) \right) \right],\label{eq::sys_sis_dt_x}\\
 w_j(t+1)&= w_j(t)+h\left(-\delta_{j}^{w}w_{j}(t)+\sum_{k=1}^{n}c_{jk}^{w}x_{k}(t)\right),\label{eq::sys_sis_dt_w}
\end{align}}
where $w_{j}$ represents the pathogen concentration in the $j$th resource; $\delta_{j}^{w}$ is the decay rate of the pathogen; $c_{jk}^{w}$ is the effective person-resource contact rate of agent {$k$} with the resource {$j$}; {$\beta_{ij}^{w}$ is the effective resource-person infection rate of the resource $j$ to the agent $i$}. In contrast to \cite{liu2019networked}, we do not assume that only one resource is shared among groups of people and our model allows multiple resources. We assume there is no direct interaction between resources, because this paper mainly focuses on studying pathogen diffusing over different types media. The inner-interaction of pathogen among the same type of medium is beyond the scope of this paper, and is left for future work.

%\hjk{we should notice here that there is no interaction between different resources and mention why}\csx{how to mention this point?}

We can simplify our discrete-time model from \eqref{eq::sys_sis_dt_x} and \eqref{eq::sys_sis_dt_w} into a matrix form as follows:
\begin{align}
    x(t+1)&= x(t)+h\left( (B-X(t)B-D)x(t) \right.\notag \\
    &\left.+(I-X(t))B_{w}w(t) \right),\label{eq::sys_sis_dt_xm}\\
    w(t+1)&= w(t)+h\left(  -D_{w}w(t)+C_{w}x(t)   \right), \label{eq::sys_sis_dt_wm}
\end{align}
where $x(t)=[x_{1}(t), \dots, x_{n}(t)]^{\top}$, $X(t)=\Dg(x(t))$, $w(t)=[w_{1}(t), \dots, w_{m}(t)]^{\top}$, $B=[\beta_{ij}]_{n \times n}$, $B_w=[\beta^w_{ij}]_{n \times m}$, $D=\Dg(\delta_{1},\dots, \delta_{n})$, $D_{w}=\Dg(\delta_{1}^{w},\dots, \delta_{n}^{w})$ and $C_{w}=[c^{w}_{jk}]_{m \times n}$. We can further use the following notations:
\begin{equation}
\label{eq::notation}
    \begin{split}
       z(t)&:=[x(t)^{\top},w(t)^{\top}]^{\top}, \;\;
        Z(t):=\left[ 
        \begin{matrix}
	\Dg(x(t)) & \mathbf{0}  \\
	\mathbf{0} & \mathbf{0}  \\
	\end{matrix}
        \right],\\
    B_{f}&:=\left[ 
        \begin{matrix}
	B & B_{w}  \\
	C_{w} & \mathbf{0}  \\
	\end{matrix}
        \right], \;
        D_{f}:=\left[ 
        \begin{matrix}
	D & \mathbf{0}  \\
	\mathbf{0} & D_{w}  \\
	\end{matrix}
        \right],\\
    \end{split}
\end{equation}

to rewrite (\ref{eq::sys_sis_dt_xm}) and (\ref{eq::sys_sis_dt_wm}) as :
\begin{equation} \label{eq::sys_sis_dt_z}
 z(t+1)= z(t)-h\left( D_{f}-(I-Z(t))B_{f}\right)z(t).
\end{equation} 

%\hjk{The way you write $z$ is wrong: by definition, $x$ and $w$ are column vectors, therefore $[x,w]$ would denote a matrix of two columns, making $z$ a vector of two rows... probably you mean $z(t)=\begin{bmatrix} x(t) \\ w(t) \end{bmatrix}$.  }\csx{i have corrected it}

%%%%%%%%%%%%%%%\begin{figure}
%%%%%%%%%%%%%%    \centering
 %%%%%%%%%%%%%   \begin{tikzpicture}[->,>=stealth',shorten >=1pt,auto,node distance=3.7cm,semithick]
 %%%%%%%%%%%% \tikzstyle{state1}=[circle,fill=yellow,draw=none,text=white]
 %%%%%%%%%%% \tikzstyle{state2}=[circle,fill=orange,draw=none,text=white]
 %%%%%%%%%% \tikzstyle{state3}=[circle,fill=green,draw=none,text=white]
 %%%%%%%%% \tikzstyle{state4}=[circle,fill=red,draw=none,text=white]
%%%%%%%%  \tikzstyle{state5}=[rectangle,fill=red,draw=none,text=white]

%%%%%%%  \node[state1]         (A)                    {$I^1$};
 %%%%%% \node[state3]         (D) [ right of=A] {$S$};
 %%%%% \node[state5]         (E) [below right of=A]       {$W$};

 %%%% \path (A) 
 %%       edge [bend left] node {$\delta_i^1$} (D)
 %%%       edge [dashed,bend right] node {} (E)
%       (D) edge [bend left]  node {$\sum \beta^1_{ij}$} (A);
 %   \path (E) edge   [dashed]       node {} (D);

%%%%%\end{tikzpicture}
 %%%%   \caption{Visualization of the model for the case of single resource. An individual
%%%is either susceptible ($S$) or infected with the virus ($I$). The resource ($W$) can be polluted by infected individuals , and in turn help the virus spread over people.}
%%\label{fig:net2}
%\end{figure}

Note that in real epidemic phenomena, the spreading rates can hardly remain time invariant. Instead, they will change over time. In order to better characterize a real spreading behavior, we modify the time-invariant model \eqref{eq::sys_sis_dt_z} into the time-varying version. Element-wise, the dynamics now read as
\begin{align}
    x_i(t+1)&= x_i(t)+h\left[ (1-x_i(t)) \left(\sum_{j=1}^{n} \beta_{ij}(t) x_j(t)\right.\right. \notag \\
&\left.\left. +\sum_{j=1}^{m} \beta_{ij}^{w}(t) w_j(t)\right) -\delta_i(t) x_i(t) \right],\label{eq::sys_sis_dt_x_t}\\
 w_j(t+1)&= w_j(t)+h\left(-\delta_{j}^{w}(t)w_{j}(t)+\sum_{k=1}^{n}c_{jk}^{w}(t)x_{k}(t)\right).\label{eq::sys_sis_dt_w_t}
\end{align}
Note that the spreading parameters $\beta_{ij}(t), \delta_i(t), \delta_{j}^{w}(t), c_{jk}^{w}(t)$ are now a function of the time $t$.

Similar to what we did above, one can further simplify the system \eqref{eq::sys_sis_dt_x_t}-\eqref{eq::sys_sis_dt_w_t} into a matrix, obtaining:
\begin{align}
    x(t+1)&= x(t)+h( (B(t)-X(t)B(t)-D(t))x(t) \notag \\
    &+(I-X(t))B_{w}(t)w(t) ),\label{eq::sys_sis_dt_xm_t}\\
    w(t+1)&= w(t)+h\left(  -D_{w}(t)w(t)+C_{w}(t)x(t)   \right), \label{eq::sys_sis_dt_wm_t}
\end{align}
where the definition of all matrices of spreading parameter $B(t),D(t),D_{w}(t), C_{w}(t)$ remain the same as in \eqref{eq::notation} but they can now change over time.

Finally, one can get the following system by using notation \eqref{eq::notation} and considering that all parameter matrices are time-varying:
\begin{equation} \label{eq::sys_sis_dt_z_t}
 z(t+1)= z(t)-h\left(D_{f}(t)-(I-Z(t))B_{f}(t)\right)z(t).
\end{equation} 

%\begin{remark}[The relationship with the other related models]
%Here, we illustrate the relationship between our model and some other related models. Compared with \cite{pare2018analysis}, we extend the system with a coupled pathogen dynamics evolved on different types of resources. To the best of our knowledge, this is the first discrete-time model to describe virus spreads over population network and different resources, although its continuous-time counterparts have been already studied by \cite{liu2019networked}, \cite{janson2020networked} and \cite{pare2021multi}. Notice that the real data is frequently gathered in discrete manner. This new contribution is important for application. Besides, all these works \cite{pare2018analysis,liu2019networked,janson2020networked,pare2021multi} is based on time-invariant parameter. Spreading parameters in our model can change over time.
%\end{remark}
%\hjk{I am not sure if the above remark should be included as exactly the same is already mentioned in the introduction. My suggestion is to delete it.}\csx{it seems reasonable to delete the remark}

\begin{remark}
In this paper, we only assume that the spreading parameters are time-varying. The study of time-varying topology remains a future work.
\end{remark} %\hjk{This sentence seems better suited for a remark instead of the previous paragraph}\csx{i agree}

\subsection{Technical Assumptions}
We introduce several assumptions on the parameters. %{which are necessary to ensure that the model is well-defined.}

{\begin{assumption}\label{ass:single}
To ensure that the model is well-defined, we assume the following.
\begin{itemize}
    \item For all $i\in [n]$, $x_{i}(0)\in [0,1]$.
    \item For all $t\geq0$, $i\in [n]$ and $j\in[n]$, $\delta_{i}(t)> 0$, $\beta_{ij}(t)\geq 0$ and $\beta^w_{ij}(t)\geq 0$. For all $i\in [n]$ and $j\in[m]$, $\delta_{j}^{w}(t)> 0$ and $c_{ij}^{w}(t) \geq 0$ with at least one $i$ such that $c_{ij}^{w}(t) > 0$.
    \item For all $t\geq0$, $j\in [m]$, $w_{j}(0) \geq 0$ and $w_{j}(0)$ is upper bounded by $ w_{\max}.$ Furthermore, $\frac{\sum_{k=1}^{n}c_{jk}^{w}(t)}{\delta_{j}^{w}(t)}\in [0, w_{\max}]$.
    \item The sampling period $h$ is positive. Furthermore, $h$ is sufficiently small so that $h\delta_{i}(t)\in (0,1]$, $h\delta_{j}^{w}(t)\in (0,1]$ and $h(\sum_{j=1}^{n} \beta_{ij}(t) +\sum_{j=1}^{n} \beta_{ij}^{w}(t) w_{\max})\in [0,1]$ hold for all $t\geq0$. 
\end{itemize}
\end{assumption}}

%\begin{assumption}\label{ass:para}
%For all $t\geq0$, $i\in [n]$ and $j\in[n]$, $\delta_{i}(t)> 0$, $\beta_{ij}(t)\geq 0$ and $\beta^w_{ij}(t)\geq 0$. For all $i\in [n]$ and %$j\in[m]$, $\delta_{j}^{w}(t)> 0$ and $c_{ij}^{w}(t) \geq 0$ with at least one $i$ such that $c_{ij}^{w}(t) > 0$.
%\end{assumption}

%\begin{assumption}\label{ass:wmax}
%For all $t\geq0$, $j\in [m]$, $w_{j}(0) \geq 0$ and $w_{j}(0)$ is upper bounded by $ w_{\max}.$ Furthermore, %$\frac{\sum_{k=1}^{n}c_{jk}^{w}(t)}{\delta_{j}^{w}(t)}\in [0, w_{\max}]$.
%\end{assumption}

%\begin{assumption}\label{ass:h}
%The sampling period $h$ is positive. Furthermore, $h$ is sufficiently small so that $h\delta_{i}(t)\in (0,1]$, $h\delta_{j}^{w}(t)\in (0,1]$ and $h(\sum_{j=1}^{n} \beta_{ij}(t) +\sum_{j=1}^{n} \beta_{ij}^{w}(t) w_{\max})\in [0,1]$ hold for all $t\geq0$. %\hjk{for all $t\geq0$? [besides, this should be lower-case $n$ right? or is there a reason why now you choose $N$? Also, this assumption implicitely requires that the $\beta$'s are bounded, which should probably be pointed-out somewhere]}.\csx{it is a typo}
%\end{assumption}
{Furthermore, we assume the corresponding layered network is strongly connected. }
\begin{assumption}\label{ass:bf}
The matrix $B_{f}(t)$ is irreducible.
\end{assumption}

\begin{remark} [Interpretation of the assumptions]
{The first statement in the Assumption \ref{ass:single} is very natural, since $x_{i}(0)$ denotes the infection possibility or the infected fraction of agent $i$. The second statement shows that the healing rate of people as well as the decay rate of resources must be positive and the infection rate of people as well as the person-resource contact rate must be non-negative. Such assumptions are consistent with the reality, since all these rates must be non-negative. Besides, each resource must be connected to at least one group of people. The third statement illustrates that the concentration is non-negative (usually, we adopt $\text{mg}/\text{liter}$ as the corresponding unit). The third statement is important to ensure the discrete-time model is well-defined. Unlike the continuous-time counterpart \cite{pare2018analysis,liu2019networked,janson2020networked,pare2021multi}, we assume that $w_{j}(0)$ is upper bounded by a value $w_{\max}$. Without this assumption, the discrete-time model may present some unrealistic behaviour which could lead to $x_{i}(t)>1$, making the system ill-defined. Moreover, if $h$ is sufficiently small, $w_{\max}$ can be sufficiently large, which corresponds to the fact that $w_{j}$ is not limited to be bounded in the continuous-time system. As for the last statement, it is needed to make sure the system has a suitable system domain. If the sampling period $h$ is small enough, one can always find appropriate spreading parameters $\delta_{i}(t),\delta_{j}^{w}(t),\beta_{ij}(t),\beta^w_{ij},c_{jk}^{w}(t)$ and $w_{\max}$ such that the assumption holds. Assumption \ref{ass:bf} illustrates that the matrix $B_{f}(t)$ is irreducible, which implies the whole equivalent graph is strongly connected. The assumption is natural due to the fact of globalization. Assumption \ref{ass:bf} is necessary to achieve analytical results and is widely adopted in related works, for instance \cite{liu2019networked,janson2020networked,pare2021multi,gracy2021analysis}.
All these assumptions \ref{ass:single}-\ref{ass:bf} are based on the time-varying system \eqref{eq::sys_sis_dt_z_t} and they also hold for the time-invariant model (7) with corresponding fixed parameters.}
\end{remark}

\begin{lemma} \label{lemma:bound}
Suppose that Assumption \ref{ass:single} hold true, for the time invariant model \eqref{eq::sys_sis_dt_z}, then $x_{i}(t) \in [0,1]$ for all $i \in [n]$ and $w_{j}(t) \in [0,w_{\max}]$ for all $j \in [m]$, for all $t\geq 0$.
\end{lemma}

\begin{pf}
See Appendix \ref{app:1}.
\end{pf}

% \begin{remark}[System domain]
% As mentioned, it is very natural to assume $x_{i}(0) \in [0,1]$. Furthermore, we see from Lemma \ref{lemma:bound} that $x_{i}(t) \in [0,1]$ for all $t\geq0$, which shows that our model is well-defined. By assumptions \ref{ass:xini}, \ref{ass:para}, \ref{ass:wmax} and \ref{ass:h}, we can ensure that $\mathbf{D}=\{ z(t)=[x(t)^\top,w(t)^\top]^{\top}\,|\,x(t)\in [0,1]^{n}, \quad  w(t)\in [0,w_{\max}]^{m}\}$ is a positively invariant set of the system given by \eqref{eq::sys_sis_dt_z}. Thus, $\mathbf{D}$ is our system domain.
% \end{remark}

\begin{remark}[System Domain] It follows from Lemma \ref{lemma:bound} that $\mathcal{D}=\{ z(t)=[x(t)^\top,w(t)^\top]^{\top}\,|\,x(t)\in [0,1]^{n}, \,  w(t)\in [0,w_{\max}]^{m}\}$ is a positively invariant set for \eqref{eq::sys_sis_dt_z}. Therefore, $\mathcal{D}$ is the system domain.%\hjk{I shortened this remark, but commented the previous version, you can decide which one you want to keep.}
\end{remark}

Now, we consider the time-varying system \eqref{eq::sys_sis_dt_z_t}.

% \begin{lemma} \label{lem:boundt}
% If $0<B_f(t)<B_{f\max}$ and $D_{f\min}<D_f(t)\leq \mathbf{1}$ such that the Assumptions 2-6 \hjk{5? or is there an assumption really missing?}are fulfilled with time invariant spreading parameters $B_{f\max}$ and $D_{f\min}$, from an initial condition under Assumption 1, for the time-varying model \eqref{eq::sys_sis_dt_z_t} with time-varying spreading parameters $B_{f}(t)$ and $D_{f}(t)$, i.e. $z(t+1)= z(t)+h\left( -D_{f}(t)+(I-Z(t))B_{f}(t)\right)z(t)$, we have $x(t) \in [0,1]$ for all $t \geq 0$ and $w(t) \in [0,w_{\max}]$ for all $t \geq 0$, where $w_{\max}$ is determined by $B_{f\max}$ and $D_{f\min}$.
% \end{lemma}

\begin{lemma}\label{lem:boundt}
Consider \eqref{eq::sys_sis_dt_z_t} satisfying Assumptions \ref{ass:single}. Suppose that there exist constant matrices $B_{f\max}$ and $D_{f\min}$ such that \eqref{eq::sys_sis_dt_z} satisfies the same assumptions \ref{ass:single} with $B_f=B_{f\max}$ and $D_f=D_{f\min}$. Moreover, let $0<B_f(t)<B_{f\max}$ and $D_{f\min}<D_f(t)\leq \mathbf{1}$. Then the solution $z(t)=[x(t)^\top,w(t)^\top]^{\top}$ of \eqref{eq::sys_sis_dt_z_t} satisfies: $x(t) \in [0,1]$  and $w(t) \in [0,w_{\max}]$ for all $t \geq 0$, where $w_{\max}$ is determined by $B_{f\max}$ and $D_{f\min}$. %\hjk{I have commented the previous version of this lemma, you can decide which one to keep.}
\end{lemma}

\begin{pf}
See Appendix \ref{app:2}.
\end{pf}

Notice that Lemma \ref{lem:boundt} implies that under the appropriate bounds of $B_f(t)$ and of $D_f(t)$, the system domain of \eqref{eq::sys_sis_dt_z_t} is also $\mathcal{D}$. {Lemmas \ref{lemma:bound} and \ref{lem:boundt} also show that all assumptions in Assumption \ref{ass:single} are natural and necessary to make sure the system is well-defined.
}
\section{Main Results on the single-virus model} \label{sec:mrs}
Before we present the properties of the time-varying model \eqref{eq::sys_sis_dt_z_t}, we need to firstly study the time-invariant counterpart \eqref{eq::sys_sis_dt_z}.
\subsection{Healthy State Behaviour}
For the system \eqref{eq::sys_sis_dt_z}, it is evident that the system has a zero equilibrium which is the so-called healthy state. In the following, we present the properties of such a healthy state and also show the existence of the so-called \textit{endemic equilibrium} with the form $z^* \gg 0$.
 \begin{thm}\label{thm:hss}
 Suppose that Assumptions \ref{ass:single}-\ref{ass:bf} hold for \eqref{eq::sys_sis_dt_z}. Then the following statements hold:
 \begin{itemize}[leftmargin=*]
     \item [i)]If $s_{1}(I-hD_f + hB_f) \leq 1$, then the healthy state is asymptotically stable with the domain of attraction $\mathcal{D}$.
     \item [ii)]If $s_{1}(I-hD_f + hB_f) > 1$, then (7) has two equilibria, 0 and $z^* \gg 0$.
     \item [iii)]The healthy state is the unique equilibrium in the domain $\mathcal{D}$ if and only if $s_{1}(I-hD_f + hB_f) \leq 1$.
 \end{itemize}
 \end{thm}
 
 \begin{pf}
See Appendix \ref{app:3}.
 \end{pf}
 
\subsection{Endemic Behaviour}
Next, we show the system endemic behaviour. It is now convenient to rewrite the system dynamics \eqref{eq::sys_sis_dt_x} and \eqref{eq::sys_sis_dt_w} into the form of error dynamics by denoting $e_{i}(t)=x_{i}(t)-x_{i}^*$ and $f_{i}(t)=w_{i}(t)-w_{i}^*$.

\begin{prop} \label{prop:2}
If the endemic equilibrium $z^*=[x^*,w^*]^{\top} \gg 0$ of the system \eqref{eq::sys_sis_dt_z} exists, the system equations \eqref{eq::sys_sis_dt_xm} and \eqref{eq::sys_sis_dt_wm} are equivalent to the error dynamics
in the matrix form, \begin{equation} \label{eq:errormatrix}
    \hat{e}(t+1)=\phi(x(t))\hat{e}(t),
\end{equation}
where \begin{equation}
\hat{e}(t)=[e_{1}(t),\dots,e_{n}(t),f_1(t),\dots,f_m(t)]^{\top} 
\end{equation}
and
\begin{equation} \label{eq:phi}
\begin{split}
    &\phi(x(t))= \\
    &\left[ 
        \begin{matrix}
	I-\Dg(\frac{h\delta_i}{1-x_i^*})+\Dg(1-x(t))hB & \Dg(1-x(t))hB_w  \\
	hC_w & I-hD_{w}  \\
	\end{matrix}\right].
\end{split}
\end{equation}
\end{prop}

\begin{pf}
See Appendix \ref{app:5}.
\end{pf}

To derive the global stability of the endemic equilibrium, we introduce a new assumption.

\begin{assumption} \label{ass:h1}
The equation $h(\delta_{i}+\sum_{j=1}^{n} \beta_{ij} +\sum_{j=1}^{m} \beta_{ij}^{w} w_{\max})\in [0,1]$ holds for all $i\in[n]$.
\end{assumption}

\begin{remark} 
Assumption \ref{ass:h1} is stricter than $h(\sum_{j=1}^{N} \beta_{ij} +\sum_{j=1}^{n} \beta_{ij}^{w} w_{\max})\in [0,1]$, which is the time-invariant version of the fourth statement in Assumption \ref{ass:single}. In other words $h(\sum_{j=1}^{N} \beta_{ij} +\sum_{j=1}^{n} \beta_{ij}^{w} w_{\max})\in [0,1]$ holds whenever Assumption 6 holds. 
\end{remark}

Now, we can confirm the global stability of the endemic equilibrium.

\begin{thm} \label{thm:end}
 Suppose that Assumptions \ref{ass:single}-\ref{ass:h1} hold for \eqref{eq::sys_sis_dt_z}. If $s_{1}(I-hD_f + hB_f) > 1$, then the endemic state is asymptotically stable with domain of attraction $\mathcal{D}\setminus \{\mathbf{0}\}$. The healthy state is unstable.
\end{thm}

\begin{pf}
See Appendix \ref{app:7}.
 \end{pf}
 
{The results for the single-virus time-invariant system we have achieved in this section are analogous to the continuous-time counterpart \cite{liu2019networked}. Regarding the endemic behaviour, we require the slightly stricter assumption \ref{ass:h1}. The study of the endemic equilibrium for the discrete-time system is usually accompanied with stricter assumptions \cite{liu2020stability,prasse2019viral}. The results of the Theorems \ref{thm:hss} and \ref{thm:end} show that our discrete-time indeed keeps the realistic properties similar to the continuous-time counterpart and succeeds in capturing the qualitative behaviour of a viral spread.}

\subsection{Reproduction number}

In epidemiology, the reproduction number, usually denoted by $R_0$, is defined to be the average number of people getting infected by one infected person. If $R_0 > 1$ the disease spreads among the population and leads to a pandemic; if $R_0 \leq 1$ the disease will gradually die out. In this section, we define the basic reproduction number of our proposed discrete-time layered  networked SIWS  model.

We have shown that  \eqref{eq::sys_sis_dt_z} converges to the healthy state if and only if $s_{1}(I-hD_f + hB_f) \leq 1$ under Assumptions 1-6. Otherwise, the system converges to the endemic equilibrium. We can rewrite the condition $s_{1}(I-hD_f + hB_f) \leq 1$ as $s_{1}(hB_f-hD_f)\leq0$. From Lemma \ref{lemma:trans}, it can be further reformulated as $\rho(D_f^{-1}B_f)\leq1$. Correspondingly, the condition of disease outbreak $s_{1}(I-hD_f + hB_f) > 1$ is equivalent to $\rho(D_f^{-1}B_f)>1$. For the conventional networked SIS model without coupled pathogen dynamics \cite{pare2018analysis}, the basic reproduction number is defined by $\rho(D^{-1}B)$. Similarly, we define $\rho(D_f^{-1}B_f)$ as the basic reproduction number of our model \eqref{eq::sys_sis_dt_z}. Notice that $\rho(D_f^{-1}B_f)$ is not well-defined if there is any node $i$ such that $\delta_{i}=0$ or $\delta_{i}^w=0$. For this reason, we constrain all $\delta_{i}$ and $\delta_{i}^w$ to be strictly positive.

Now, we show the relationship of the basic reproduction number between both models.

\begin{prop}\label{prop:comp}
Suppose that Assumptions \ref{ass:single}-\ref{ass:h1} hold. The
basic reproduction number of the layered networked SIWS
model \eqref{eq::sys_sis_dt_z} is greater than that of the networked SIS model \cite{pare2018analysis}.
\end{prop}

\begin{pf}
The proposition is derived by observing that $D^{-1}B$ is the principal square submatrix of $D_{f}^{-1}B_f$ and applying Lemma \ref{lemma:comp} in the appendix \ref{app:prononneg}. 
\end{pf}

To better understand the reproduction number, we provide the following results.

\begin{prop}\label{prop:crate}
Suppose that Assumptions \ref{ass:single}-\ref{ass:bf} hold for \eqref{eq::sys_sis_dt_z}. If the basic reproduction number $\rho(D_f^{-1}B_f)\leq 1$, then the healthy state is locally exponentially stable with a convergence rate given by $s_1(I-hD_f+hB_f)$.
\end{prop}

\begin{pf}
See Appendix \ref{app:crate}.
\end{pf}

Proposition \ref{prop:crate} shows that the reproduction number less than or equal to $1$ directly implies the local exponential stability of the healthy state and how fast it converges locally. We can observe from the simulation in section \ref{sec:simsig} that whenever $s_1(I-hD_f+hB_f)\leq 1$, the greater $s_1(I-hD_f+hB_f)$ is, the slower the model converges to the origin. From such simulation, it seems that under Assumptions \ref{ass:single}-\ref{ass:bf} and $s_{1}(I-hD_f + hB_f) \leq 1$, the healthy state is globally exponentially stable with a convergence rate given by $s_1(I-hD_f+hB_f)$. This conjecture remains part of a future work.

Besides, the infection (contamination) level of the endemic equilibrium also depends on the reproduction number. We might as well consider the homogeneous case. We can validate that it is true when the model is homogeneous.

\begin{prop} \label{prop:homo}
Suppose that Assumptions \ref{ass:single}-\ref{ass:bf} hold for \eqref{eq::sys_sis_dt_z}. Moreover, let $\beta_{ij}=\beta$ and $\delta_i=\delta$ for all $i\in[n],\,j\in [n]$. Moreover, for all $i\in[n],j\in[m]$, let $\delta^w_j=\delta^w$, $\beta^w_{ij}=\beta$ and $c^w_{ji}=c$. Also, define $\hat{c}\coloneqq\frac{c}{\delta^w}$. The endemic equilibrium of this homogeneous system reads as $x^*=1-\frac{\delta}{n\beta(1+m\hat{c})}$ and $w^*=n\hat{c}x^*$.
\end{prop}

\begin{pf}
See Appendix \ref{app:homo}.
\end{pf}

For the conventional discrete-time SIS model \cite{pare2018analysis} under homogeneous parameter setting, the solution reduces to $x^*=1-\frac{\delta}{n\beta}$. From Proposition \ref{prop:homo}, we see that if any entry of $D_f$ (i.e, $\delta$ and $\delta^w$) decreases or any entry of $B_f$ (i.e, $c$, $\beta$ and $\beta^w$) increases, the infection level $x^*$ as well as the contamination level $w^*$ of the endemic equilibrium increase and vice versa. 

Then, we investigate the heterogeneous case under some small perturbation of $D_f$ and $B_f$. We find out that the statement is still true under this circumstance.

% \begin{prop} \label{prop:pertubation}
% Suppose that Assumptions 1-5 hold for \eqref{eq::sys_sis_dt_z}. If $s_{1}(I-hD_f + hB_f) > 1$, each entry of the epidemic state $z_i^*$ is the a strictly decreasing function of the entry of $D_f$ and is a strictly increasing function of the entry of $B_f$ around the point $z^*$.
% \end{prop}

\begin{prop} \label{prop:pertubation}
Suppose that Assumptions \ref{ass:single}-\ref{ass:bf} hold for \eqref{eq::sys_sis_dt_z}. If $s_{1}(I-hD_f + hB_f) > 1$, then each $i$-th component of the epidemic state $z^*$ is a strictly decreasing function of the $i$-th entry of the diagonal matrix $D_f$ and is a strictly increasing function of any entry of the $i$-th row of $B_f$.
\end{prop}

\begin{pf}
See Appendix \ref{app:8}.
\end{pf}

However, it still remains an open question how the endemic equilibrium will change when the perturbation is big resulting in the new endemic equilibrium far from the original one. The simulation in the section \ref{sec:simsig} suggests that the result still holds true when the perturbation is large.

\subsection{Time-varying spreading parameter}

We now focus on the time-varying model \eqref{eq::sys_sis_dt_z_t} and discover its system behaviour. Firstly, we discover the condition under which the time-varying model converge to the healthy state.

%It worth noticing that if $0<B_f(t)<B_{f\max}$ and $D_{f\min}<D_f(t)\leq \mathbf{1}$ such that the Assumptions 2-5 are fulfilled with time invariant spreading parameters $B_{f\max}$ and $D_{f\min}$, it already shows that $B_f(t)$ and $D_f(t)$ will satisfy Assumptions 2-5 for all $t\geq 0$.

% \begin{thm} \label{thm:6}
% If $0<B_f(t)<B_{f\max}$ and $D_{f\min}<D_f(t)\leq \mathbf{1}$ such that Assumptions \ref{ass:para}-\ref{ass:h1} are fulfilled with time invariant spreading parameters $B_{f\max}$ and $D_{f\min}$ and $s_{1}(I-hD_{f\min} + hB_{f\max}) \leq 1$, under Assumption \ref{ass:xini}, for the time-varying model \eqref{eq::sys_sis_dt_z_t} with time-varying spreading parameters $B_{f}(t)$ and $D_{f}(t)$, i.e. $z(t+1)= z(t)+h\left( -D_{f}(t)+(I-Z(t))B_{f}(t)\right)z(t)$, the healthy state is asymptotically stable with the domain of attraction $\mathbf{D}$.
% \end{thm}

\begin{thm}\label{thm:6}
Consider \eqref{eq::sys_sis_dt_z_t} and suppose that there exist constant matrices $B_{f\max}$ and $D_{f\min}$ such that $0<B_f(t)<B_{f\max}$ and $D_{f\min}<D_f(t)\leq \mathbf{1}$. Moreover, suppose that there hold Assumptions \ref{ass:single}-\ref{ass:bf}, and $s_{1}(I-hD_{f\min} + hB_{f\max}) \leq 1$  for $B_{f}(t)$, $D_{f}(t)$, and $B_{f\max}$, $D_{f\min}$. Then, the healthy state of \eqref{eq::sys_sis_dt_z_t} is asymptotically stable with domain of attraction $\mathbf{D}$.
\end{thm}

\begin{pf}
 See Appendix \ref{app:9}.
 \end{pf}
 
Inspired by \cite{sereno2022minimizing,bertozzi2020challenges}, we further suggest $\rho(D^{-1}_{f}(t) B_{f}(t))$ as a time-varying reproduction
number, and $\rho(D^{-1}_{f\min} B_{f\max})$ as an effective reproduction number for the time-varying case, since the latter term directly indicates whether the healthy state is globally stable and the first term is a real-time value of the spectral radius. In the section \ref{sec:sim}, we show by numerical examples that the healthy state becomes unstable once $s_{1}(I-hD_{f\min} + hB_{f\max}) \leq 1$ is violated.
 
Although the analysis of the healthy state is straightforward, the analysis of endemic behaviour when $s_{1}(I-hD_{f\min} + hB_{f\max}) > 1$ is considerably more complicated for the time-varying case. In essence, there is no guarantee that the system converges to a certain endemic equilibrium, and even the existence of the endemic equilibrium is difficult to obtain. Here, we consider the following simple example to illustrate our arguments. We consider the time-varying system \eqref{eq::sys_sis_dt_z_t} which can be treated as a switched system between 2 time-invariant system \eqref{eq::sys_sis_dt_z} with parameters $B_{f1}, D_{f1}$ and $B_{f2}, D_{f2}$. We also assume $B_{f1}, D_{f1}$ and $B_{f2}, D_{f2}$ satisfy Assumptions 2-6 and refer to the endemic equilibrium $z^{*1}$ and $z^{*2}$ respectively. We consider the switching time is sufficiently large so that the model always have time to converge before each switch. Assumption 1 holds generally. In this case, we can observe that the system will oscillate between the two endemic equilibria. If we consider the switching among $n$ time-invariant system, the result will be similar.

Even though we can not guarantee the existence of endemic equilibrium and system convergence to the endemic equilibrium in general, it is still very interesting to see under which circumstance the time-varying system will converge to a certain endemic equilibrium. We define the system parameters pair as $(B_f,D_f)$. Now, we define the set $\Omega_{z^*}$, which conclude all the pairs $(B_f,D_f)$ that refer to the same endemic equilibrium $z^*$ for the time invariant system \eqref{eq::sys_sis_dt_z} and satisfy Assumption 2-6. 

We can observe that the pair $(B_f,D_f)$ that refers to a certain endemic equilibrium $z^*$ is not unique. For the homogeneous case, this is the direct consequence of proposition \ref{prop:homo}. This makes $\Omega_{z^*}$ not empty.

Now, we consider the time-varying system \eqref{eq::sys_sis_dt_z_t}  as a switched system, where its parameter pair $(B_f(t),D_f(t))\in \Omega_{z^*}$ for all $t\geq 0$. The following theorem reveals that the system will finally converge to this endemic equilibrium $z^*$, to which the set $\Omega_{z^*}$ refers. Notice that in this case, the existence of $\Omega_{z^*}$ already shows that $s_{1}(I-hD_{f}(t) + hB_{f}(t) > 1$ holds for all $t\geq 0$, since the endemic equilibrium must exist.

\begin{thm} \label{thm:7}
For the system \eqref{eq::sys_sis_dt_z_t}, if $B_f(t)$ is symmetric for all $t\geq 0$ and $(B_f(t),D_f(t))\in \Omega_{z^*}$, where $\Omega_{z^*}$ is the set that concludes all the pairs $(B_f,D_f)$ that refer to the same endemic equilibrium $z^*$ for the time invariant system \eqref{eq::sys_sis_dt_z} and all these pairs satisfy Assumption \ref{ass:single}-\ref{ass:h1}.  The endemic equilibrium $z^*$ is asymptotically stable with domain of attraction $\mathcal{D}\setminus \{\mathbf{0}\}$.
\end{thm}

%\hjk{The above situation needs more explanation. Do you mean that $B_f(t)$ takes constant values in $\Omega$ for certain intervals of time, and then switches to another constant value that leads to the same equilibrium? If so, all this needs to be specified, does the argument depend on the switching frequency? In any case, I am not sure about the relevance of this result to be honest, we can talk about it later.} \csx{no, there is no restriction about the switch frequency. The only condition is $B_f(t)$ and $D_f(t)$ are always in the set $\Omega_{z^*}$ and $B_f(t)$ is symmetric.}

\begin{pf}
See Appendix \ref{app:10}.
\end{pf}
%\hjk{overall comment: in the previous section some statements require assumptions 1-5 and some 1-6, please make sure all ask for precisely what they need.}\csx{it looks strange but it is true. the results regarding the healthy state are based on 1-5. The results of endemic behaviour depends on 1-6. if 1-6 are satisfied, 1-5 are satified.}

{In \cite{gracy2021analysis}, a time-varying continuous-time counterpart is studied and a result of global exponential convergence to the healthy state with a slightly stronger assumption is given. In our paper, the result of global asymptotic convergence with a mild assumption is provided. To our best knowledge, no literature has studied the endemic behaviour in either continuous-time or discrete-time context for the networked SIWS or SIS model. We have solved this open question by adopting a relatively strict assumption.}

\section{Extension to the multi-virus model} \label{sec:multi}
In this section, we generalize the discrete-time networked SIWS model with multiple resources ~\eqref{eq::sys_sis_dt_z} into networked multi-virus multi-layer multi-resource SIWS model by introducing multiple competing viruses.

\subsection{Model Set-up}

Consider $l$ viruses spreading over an $l+m$ layer graph $\mathcal{G}$, where the $k$-th virus spreads via the $k$-th layer $\mathcal{G}^k$ and the last $m$ layers are the different types of resources. The first $k$ layer forms a multiplex network of human contact. The multiplex network contains  multiple systems of the same set of nodes and there exists various types of the interactions among nodes, see \cite{kinsley2020multilayer,mittal2018anomaly} for more about multiplex and multi-layer networks. In the human contact multiplex networks, the nodes in different layers represent the same population of individuals but with different forms of contact, depending on the properties of different viruses. The human contact multiplex network together with resources forms an interconnected network. The interconnected network denotes the network where each layer denotes different entities, see also \cite{kinsley2020multilayer} for more about the interconnected network.

\begin{figure}
    \centering
    \begin{tikzpicture}
    \node at (0,0){\includegraphics[scale=.95]{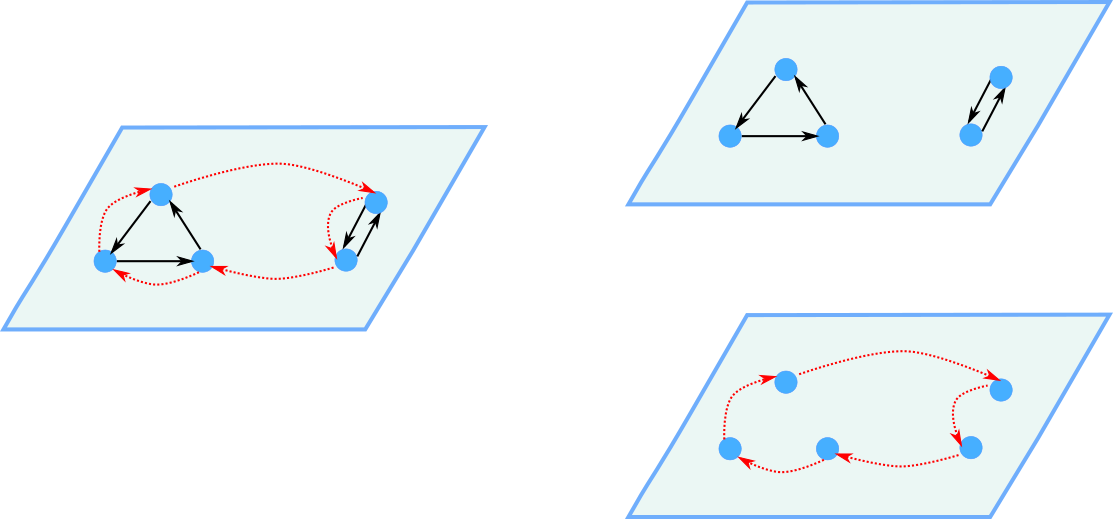}};
    \node at (0,0){\Large$=$};
    \node at (2,0){\Large$+$};
    \end{tikzpicture}
    
    \caption{An example of a 2 layer human contact multiplex network. The edges with different colors denote the different ways of interaction varying from virus to virus. The subgraph of each layer is not assumed to be strongly connected.}
    
    \label{fig:multiplex}
\end{figure}

\begin{figure}
    \centering
    \begin{tikzpicture}
    \node at (0,0){
    \includegraphics[scale=1]{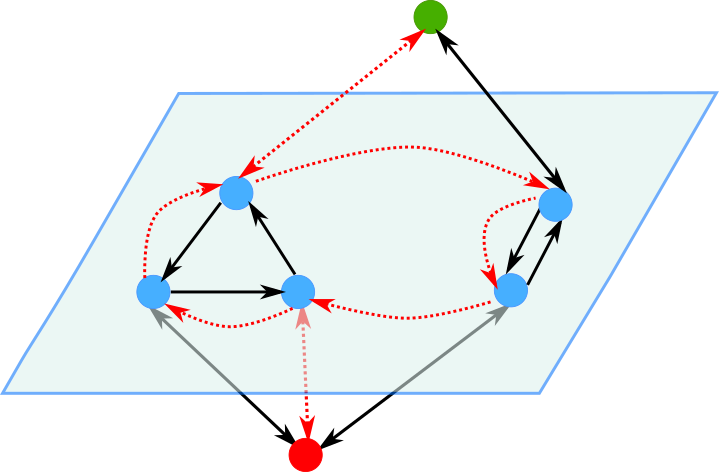}
    };
    \node at (2.2,1.8){\small{ventilation system}};
    \node at (0.2,-1.9){\small {water}};
    \node[blue!70!green] at (-2,1.4){\small {human-contact network}};
    \end{tikzpicture}
    \caption{An example of a human contact multiplex network plus two different resources. The subgraph of virus $k$ in the human contact multiplex network together with resorces forms the graph $\mathcal{G(B^{\text{k}}_{\text{f}})}$, which is assumed to be strongly connected. The way that resources interact with different layers of the human contact network may also vary. We use different colours to highlight different types of diffusion due to different kinds of viruses.
    }
    \label{fig:mulnet1}
\end{figure}
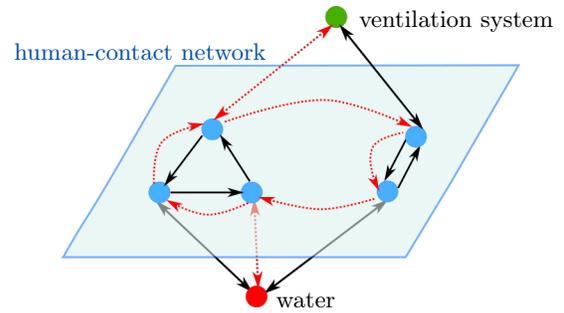

Figure \ref{fig:multiplex} shows an example of the human contact multiplex network. Figure \ref{fig:mulnet1} illustrates an example of the whole equivalent multi-virus graph structure.

One of the most important assumption for modelling is that each $k$-th agent is assumed to be infected with only one virus at a time and not multiple viruses simultaneously. Co-infection at the same time is not possible. Besides, akin to the single-virus case, the resource can be polluted by infected individuals, and in turn help the virus spread over people. However, the pathogen of different viruses can coexist in the resource.

Thus, the model we propose is given, element-wise, as follows
{\small\begin{align}
     x_i^k(t+1)&= x_i^k(t)+h\left[ \left(1-\sum^{l}_{a=1}x^a_i(t)\right) \left(\sum_{j=1}^{n} \beta^k_{ij}(t) x^k_j(t)\notag\right.\right. \\
     &\left.\left.+\sum_{j=1}^{m} \beta_{ij}^{wk}(t) w^k_j(t)\right) -\delta_i^k(t) x^k_i(t) \right], \label{eq::sys_sis_dt_xmult} \\
  w_j^k(t+1)&= w_j^k(t)+h\left(-\delta_{j}^{wk}(t)w^k_{j}(t)+\sum_{a=1}^{n}c_{ja}^{wk}(t)x^k_{a}(t)\right).\label{eq::sys_sis_dt_wmult}
\end{align}}
where all variables and parameters refer to the same meaning as the single-virus model and the subscript $k$ denotes that it is the variable or parameter of virus $k$.

%$x^k_{i}$ denotes the agent $i$'s infection probability of the $k$-th virus or the infection proportion of group $i$ to the $k$-th virus; $\beta^k_{ij} := \beta^k_{i} A^k_{ij}$, where $\beta_{i}^k$ is the infection rate of node $i$ to the $k$-th virus and $A_{ij}^k$ is the entry of the subgraph $\mathcal{G}^k$; $w_{j}^k$ represents the pathogen's concentration of the $k$-th virus in the $j$-th resource; $\delta_{j}^{wk}$ is decay rate of the $k$-th pathogen; $c_{jk}^{wk}$ is the effective $k$-th virus's person-resource contact rate per person or agent $j$ to the resource $k$.

We can rewrite equations \eqref{eq::sys_sis_dt_xmult} and \eqref{eq::sys_sis_dt_wmult} into the matrix form:
\begin{equation} \label{eq::sys_sis_dt_xm1t}
\begin{split}
     &x^k(t+1)= x^k(t)+h( (I-\sum_{a=1}^{l}\Dg(x^a(t))B^k(t)\\
     &-D^k(t))x^k(t)+(I-\sum_{a=1}^{l}\Dg(x^a(t)))B^k_{w}(t)w^k(t) ),
\end{split}
\end{equation} 
\begin{equation} \label{eq::sys_sis_dt_wm1t}
 w^k(t+1)= w^k(t)+h\left(  -D^k_{w}(t)w^k(t)+C^k_{w}(t)x^k(t)   \right);
\end{equation}
where $x^k(t)=[x^k_{1}(t), \dots, x^k_{n}(t)]^{\top}$, $w^k(t)=$ \\$ [w^k_{1}(t), \dots, w^k_{m}(t)]^{\top}$, $B^k=[\beta^k_{ij}]_{n \times n}$, $B^k_w=[\beta^w_{ij}]_{n \times m}$, $D^k=\Dg(\delta^k_{1},\dots, \delta^k_{n})$, $D_{w}^k=\Dg(\delta_{1}^{wk},\dots, \delta_{n}^{wk})$ and $C_{w}^k=[c^{wk}_{jk}]_{m \times n}$.

We can further rewrite the systems~\eqref{eq::sys_sis_dt_xm1t} and~\eqref{eq::sys_sis_dt_wm1t} as follows:
{\small\begin{equation} \label{eq::sys_sis_dt_zmult}
z^k(t+1)= z^k(t)-h\left( +D^k_{f}(t)-(I-\sum_{a=1}^{l}Z^a(t))B^k_{f}(t)\right)z^k(t).
\end{equation}}
where the matrix or vector of $z^k(t),Z^k(t),B^k_{f}(t),\\D^k_{f}(t)$ has the same structure and meaning of \eqref{eq::notation} and the subscript $k$ denotes that it is the matrix or vector of virus $k$.

%\begin{equation}
%\label{eq::notationmul}
%    \begin{split}
%       z^k(t):=[x^k(t)^\top,w^k(t)^\top]^{\top}, \\
%        Z^k(t):=\left[ 
%        \begin{matrix}
%	\Dg(x^k(t)) & \mathbf{0}  \\
%	\mathbf{0} & \mathbf{0}  \\
%	\end{matrix}
%       \right],\\
%    B^k_{f}:=\left[ 
%        \begin{matrix}
%	B^k & B^k_{w}  \\
%	C^k_{w} & \mathbf{0}  \\
%	\end{matrix}
%        \right],\\
%        D^k_{f}:=\left[ 
%        \begin{matrix}
%	D^k & \mathbf{0}  \\
%	\mathbf{0} & D^k_{w}  \\
%	\end{matrix}
%        \right].\\
%    \end{split}
%\end{equation}

%\begin{remark}
%Consider the model \eqref{eq::sys_sis_dt_zmul} in the case of only one shared resource ($m=1$). Such a model can also be derived by applying Euler's Method directly to the continuous-time system in \cite{janson2020networked}. In the case of no resource involved, the model reduces to the discrete-time competing virus model introduced in \cite{pare2020analysis}.
%\end{remark}

For the time-invariant multi-virus system, the parameters is fixed and $(t)$ behind the parameters can be omitted. Readers can refer to appendix \ref{app:11} for precise formulations of the time-invariant model.

%\hjk{I think that for the multi-virus multi-resource model, showing again the time-invariant and time-varying models is too repetitive. We could just show the time-varying one as the use of the time invariant version is already clear from the previous section, and this could save some space.}\csx{I don't think it will save the space because even if we do that, we should also list system equations for time invariant case since they are so important. in fact, most results are based on time invariant case.}
%\csx{maybe rephrase this section later to save space}

\subsection{Technical assumptions}

Here, we propose multi-virus version of our technical assumptions, which are similar to the single-virus version \ref{ass:single}-\ref{ass:h1}.

{\begin{assumptionp}{\ref{ass:single}b} \label{ass:multi}
To ensure the model is well-defined, we assume the following.
\begin{itemize}
    \item For all $i\in [n]$, $\sum_{k=1}^{l} x^k_{i}(0)\in [0,1]$.
    \item For all $t\geq0$, $i\in [n]$, $k\in [l]$ and $j\in[n]$, $\delta^k_{i}(t)> 0$, $\beta^k_{ij}(t)\geq 0$ and $\beta^{wk}_{ij}(t)\geq 0$. For all $k\in [l]$, $i\in [n]$ and $j\in[m]$, $\delta^{kw}_{j}(t)> 0$ and $c_{ij}^{wk}(t) \geq 0$ with at least one $i$ such that $c_{ij}^{wk} > 0$.
    \item For all $t\geq0$, $k\in [l]$, $i\in [n]$ and $j\in [m]$, $w^k_{j}(0) \geq 0$ and $w^k_{j}(0)$ is upper bounded by $ w^k_{\max}.$ Furthermore, $\frac{\sum_{k=1}^{n}c_{jk}^{wk}(t)}{\delta_{j}^{wk}(t)}\in [0, w^k_{\max}]$.
    \item The sampling period $h$ is positive. Furthermore, for all $t\geq0$, $k\in [l]$, $h\delta_{i}^k(t)\in (0,1]$, $h\delta_{j}^{wk}(t)\in (0,1]$ and $h\sum_{k=1}^{l}(\sum_{j=1}^{n} \beta^k_{ij}(t) +\sum_{j=1}^{n} \beta_{ij}^{wk}(t) w^k_{\max})\in [0,1]$.
\end{itemize}
\end{assumptionp}}

%\begin{assumptionp}{\ref{ass:para}b} \label{ass:param}
%For all $t\geq0$, $i\in [n]$, $k\in [l]$ and $j\in[n]$, $\delta^k_{i}(t)> 0$, $\beta^k_{ij}(t)\geq 0$ and $\beta^{wk}_{ij}(t)\geq 0$. For all %$k\in [l]$, $i\in [n]$ and $j\in[m]$, $\delta^{kw}_{j}(t)> 0$ and $c_{ij}^{wk}(t) \geq 0$ with at least one $i$ such that $c_{ij}^{wk} > 0$.
%\end{assumptionp}

%\begin{assumptionp}{\ref{ass:wmax}b} \label{ass:wmaxm}
%For all $t\geq0$, $k\in [l]$, $i\in [n]$ and $j\in [m]$, $w^k_{j}(0) \geq 0$ and $w^k_{j}(0)$ is upper bounded by $ w^k_{\max}.$ Furthermore, $\frac{\sum_{k=1}^{n}c_{jk}^{wk}(t)}{\delta_{j}^{wk}(t)}\in [0, w^k_{\max}]$.
%\end{assumptionp}

%\begin{assumptionp}{\ref{ass:h}b} \label{ass:hm}
%The sampling period $h$ is positive. Furthermore, for all $t\geq0$, $k\in [l]$, $h\delta_{i}^k(t)\in (0,1]$, $h\delta_{j}^{wk}(t)\in (0,1]$ and $h\sum_{k=1}^{l}(\sum_{j=1}^{n} \beta^k_{ij}(t) +\sum_{j=1}^{n} \beta_{ij}^{wk}(t) w^k_{\max})\in [0,1]$.
%\end{assumptionp}

\begin{assumptionp}{\ref{ass:bf}b} \label{ass:bfm}
The matrix $B_{f}^k(t)$ is irreducible for all $k\in [l]$ and $t\geq0$,.
\end{assumptionp}

\begin{assumptionp}{\ref{ass:h1}b} \label{ass:h1m}
The equation $h(\delta^k_{i}(t)+\sum_{j=1}^{N} \beta^k_{ij}(t) +\sum_{j=1}^{N} \beta_{ij}^{wk}(t) w^k_{\max})\in [0,1]$ holds for all $k\in [l]$ and $t\geq0$.
\end{assumptionp}

All Assumptions \ref{ass:multi}-\ref{ass:h1m} are defined based on the time-varying model \eqref{eq::sys_sis_dt_zmult}. They are still valid for the time-invariant model, where the parameters are fixed. They are the multi-virus version of assumptions  \ref{ass:single}-\ref{ass:h1}. {The assumptions \ref{ass:multi} are necessary to make sure the model is well-defined. The assumptions \ref{ass:bfm}-\ref{ass:h1m} are necessary to achieve analytical results.}

The following Lemma shows that the time-invariant multi-virus model is upper bounded by its single-virus counterpart.

\begin{lemma}\label{lemma:boundmul}
Suppose that Assumptions \ref{ass:multi} hold, from any initial condition such that $x^k_{i}(0)\in [0,1]$ and $w^k_{j}(0)\in [0,w^k_{\max}]$ for any $i\in [n]$ and $j\in[m]$, then the $k$-th virus dynamic 
\begin{equation*}
 z^k(t+1)= z^k(t)+h\left( -D^k_{f}+(I-\sum_{a=1}^{l}Z^a(t))B^k_{f}\right)z^k(t). 
\end{equation*} 
is upper bounded by its single-virus counterpart 
\begin{equation*}
 \hat{z}^k(t+1)= \hat{z}^k(t)+h\left( -D^k_{f}+(I-\hat{Z}(t))B^k_{f}\right)\hat{z}^k(t),
\end{equation*} 
which originates from the same initial condition, i.e. if $z^k(0)=\hat{z}^k(0)$ then $z^k(t)\leq \hat{z}^k(t), \forall t \geq 0$ .
\end{lemma}

\begin{pf}
See Appendix \ref{app:11}.
\end{pf}

\begin{remark} 
The single-virus counterpart can be seen as the $z^k(t)$ when other viruses die out, i.e. $z^a(t)=0$ for all $t\geq0$ and $a \neq k.$
\end{remark}

\begin{lemma}\label{lemma:inimul}
Suppose that assumption \ref{ass:multi} hold true, then $x_{i}^k(t) \in [0,1]$, $w^k_{j}(t) \in [0,w^k_{\max}]$,  and $\sum_{a=1}^{l}x_{i}^a(t) \in [0,1]$ for all $k\in[l]$, $t\geq 0$, $i \in [n]$ and $j \in [m]$.
\end{lemma}

\begin{pf}
See Appendix \ref{app:12}.
\end{pf}

\begin{remark}[System domain]
Lemma \ref{lemma:inimul} also shows that the multi-virus model is well-defined. Define $\mathcal{D^\text{k}}=\{ z^k(t)=(x^k(t),w^k(t))^{\top}|x^k(t)\in [0,1]^{n}, \quad  w^k(t)\in [0,w^k_{\max}]^{m}, \text{for any $i\in [n]$ }\}$. The system domain is then $\mathcal{E}=\{A(t)=[Z^1(t),\dots,Z^l(t)]^{\top}| z^k(t)\in\mathcal{D^\text{k}},$ for any $k\in [l]$ and $\sum_{a=1}^{l}x_{i}^a(t)\in [0,1], \forall t \geq 0\}$.
\end{remark}

Now, we consider the time-varying system \eqref{eq::sys_sis_dt_zmult}.

\begin{lemma}\label{lemma:inimul1}
Consider the system \eqref{eq::sys_sis_dt_zmult} and suppose that there exists constant matrices $B^k_{f\max}$ and $D^k_{f\min}$ such that $0<B^k_f(t)<B^k_{f\max}$ and $D^k_{f\min}<D^k_f(t)\leq \mathbf{1}$ hold for every $k\in [l]$. Moreover, suppose that Assumptions \ref{ass:multi} hold for any $B^k_{f}(t)$ and $D^k_{f}(t)$ and $B^k_{f\max}$, $D^k_{f\min}$, we have $x_{i}^k(t) \in [0,1]$, $w^k_{j}(t) \in [0,w^k_{\max}]$,  and $\sum_{a=1}^{l}x_{i}^a(t) \in [0,1]$ for all $a\in[l]$, $t\geq 0$, $i \in [n]$ and $j \in [m]$, where $w^k_{\max}$ is determined by $B^k_{f\max}$ and $D^k_{f\min}$.
%If $0<B^k_f(t)<B^k_{f\max}$ and $D^k_{f\min}<D^k_f(t)\leq \mathbf{1}$ such that the Assumptions \ref{ass:param}-\ref{ass:h1m} are fulfilled with time invariant spreading parameters $B^k_{f\max}$ and $D^k_{f\min}$ for all $k\in [l]$, from an initial condition under Assumption \ref{ass:xinim}, for the time-varying model \eqref{eq::sys_sis_dt_zmult} with time-varying spreading parameters $B_{f}(t)$ and $D_{f}(t)$, we have $x_{i}^k(t) \in [0,1]$, $w^k_{j}(t) \in [0,w^k_{\max}]$,  and $\sum_{a=1}^{l}x_{i}^a(t) \in [0,1]$ for all $a\in[l]$, $t\geq 0$, $i \in [n]$ and $j \in [m]$, where $w_{\max}$ is determined by $B_{f\max}$ and $D_{f\min}$. \hjk{I think we should rephrase this lemma as in the single virus}\csx{sure, i will do that}
\end{lemma}

\begin{pf}
See Appendix \ref{app:13}.
\end{pf}

Thus, the system domain of \eqref{eq::sys_sis_dt_z_t} is also $\mathcal{E}$. Moreover, just as for fixed parameters, we have the following:

\begin{lemma}\label{lemma:boundmul1}
The $k$-th virus time-varying dynamic 
{\small\begin{equation*}
 z^k(t+1)= z^k(t)+h\left( -D^k_{f}(t)+(I-\sum_{a=1}^{l}Z^a(t))B^k_{f}(t)\right)z^k(t). 
\end{equation*}} 
is upper bounded by its single-virus counterpart
\begin{equation*}
 \hat{z}^k(t+1)= \hat{z}^k(t)+h\left( -D^k_{f}(t)+(I-\hat{Z}(t))B^k_{f}(t)\right)\hat{z}^k(t),
\end{equation*} 
which originates from the same initial condition, i.e. $z^k(t)\leq \hat{z}^k(t), \forall t \geq 0$.
\end{lemma}

\begin{pf}
See Appendix \ref{app:14}.
\end{pf}

\subsection{Analytical Results of Time-invariant Case }
Based on the time invariant system \eqref{eq::sys_sis_dt_zmul}, we provide the following analytical results. Firstly, we discover under which situation all viruses will finally die out and accordingly the healthy state will be achieved.

\begin{thm} \label{thm:hmul}
Suppose that Assumptions \ref{ass:multi}-{\ref{ass:bfm}} hold. If $s_1(I-D_f^k + B_f^k
) \leq 1$ for all $k\in [l]$, then the healthy
state is the unique equilibrium of system \eqref{eq::sys_sis_dt_zmul}, which is asymptotically stable with domain of attraction $\mathcal{E}$.
\end{thm}

\begin{pf}
See Appendix \ref{app:15}.
\end{pf}

Sometimes, one virus will dominate the spreading process and leads to a 'winner takes all' situation. That means only one virus will survive after competition and all other will die out. We call this kind of equilibrium dominant endemic equilibrium. The following theorem confirms convergence to dominant endemic equilibrium when there is only one virus with reproduction number greater than $1$.

\begin{thm} \label{thm:mul1}
Suppose that Assumptions \ref{ass:multi}-\ref{ass:h1m} hold. If $s_1(I-D_f^k + B_f^k
) > 1$ for a certain $k$ and $s_1(I-D_f^a + B_f^a
) \leq 1$ for all $a\neq k$, then the system \eqref{eq::sys_sis_dt_zmul} has two equilibra. The healthy state is asymptotically stable with domain of attraction $\mathcal{H}=\{A(t)=[Z^1(t),\dots,Z^l(t)]^{\top}\,|\, z^a(t)\in\mathcal{D^\text{k}} \text{ for all $a\in [l]$ with $a\neq k$, and } z^k(t)=\mathbf{0}\}$. The unique dominant endemic equilibrium $(\mathbf{0},\dots,z^{k*},\dots,\mathbf{0})$ with $z^{k*} \gg \mathbf{0}$ is asymptotically stable with domain of attraction $\mathcal{E}\setminus \mathcal{H}$.
\end{thm}

\begin{pf}
See Appendix \ref{app:16}.
\end{pf}

{Theorem \ref{thm:mul1} is similar to its continuous-time counterpart \cite{janson2020networked}. However, in \cite{janson2020networked} only one resource is involved. On the other hand, in \cite{pare2020analysis}, the existence of dominant endemic equilibrium and its local stability is provided. However, notice that if we impose all the pathogen to be zero in the resource, our model recovers the discrete-time SIS mode. Moreover, Theorem \ref{thm:mul1} extends the local stability result into a global one.}

Next, to further our insight, we focus on the scenario of two competing virus with a shared resource, i.e. $l=2$ and $m=1$. In this case, our system is equivalent to the discrete-time system derived by applying Euler's Method to the continuous-time model \cite{janson2020networked}. 

% \begin{pf}
% Let the Jacobian matrix of the corresponding discretized system $x(t+1)=x(t)+hf(x(t))$ at $x^*$ be denoted by $J^2(x^*)$. Thus, $J^2(x^*)=hJ^1(x^*)+I$, which implies that an eigenvalue $\lambda^2$ of $J^2(x^*)$ is given by $\lambda^2=h\lambda^1+1$ where $\lambda^1$ is an eigenvalue of $J^1(x^*)$. Next, it follows that if $\Re(\lambda^1)>0$ then $|\lambda^2|>1$ showing the first item of the statement.
% \end{pf}

Before we describe the coexistence equilibrium, let us consider the following cases: if $s_1(I-hD_f^1 + hB_f^1) > 1$ and $s_1(I-hD_f^2 + hB_f^2) \leq 1$, system \eqref{eq::sys_sis_dt_zmul} converges to 
\begin{equation} \label{eq:z*1}
(z^{*1},\mathbf{0}). 
\end{equation}
Conversely, if $s_1(I-hD_f^1 + hB_f^1) \leq 1$ and $s_1(I-hD_f^2 + hB_f^2) > 1$, system \eqref{eq::sys_sis_dt_zmul} converges to
\begin{equation} \label{eq:z*2}
    (\mathbf{0},z^{*2}).
\end{equation}
The above is the direct consequences of theorem \ref{thm:mul1}. With this in mind, let us define the notation 
\begin{equation}
\label{eq::notation1}
    \begin{split}
        Z^{*k}=\left[ 
        \begin{matrix}
	\Dg(x^{*k}) & \mathbf{0}  \\
	\mathbf{0} & \mathbf{0}  \\
	\end{matrix}
        \right],
        \end{split}
\end{equation}
for $k \in [2]$. {Following \cite{janson2020networked}, we give the next results.} {Firstly, }the following {proposition} reveals the existence of a coexisting equilibrium. Different from dominant endemic equilibrium, it shows the coexistence of viruses in such a equilibrium.

% \begin{thm} \label{thm:coexist}
% Under Assumptions \ref{ass:xinim}-\ref{ass:h1m}, with $l=2$ and $m=1$, for the model \eqref{eq::sys_sis_dt_zmul}, if $s_1(I-hD_f^k + hB_f^k) > 1$ for $k=1,2$ and $s_1(-hD_f^1 +h(I-Z^{*2}) B_f^1) > 0$ and $s_1(-hD_f^2 +h(I-Z^{*1}) B_f^2) > 0$ with $z^{*1}$, $z^{*2}$ and $Z^{*1}$, $Z^{*2}$ defined as in \eqref{eq:z*1}, \eqref{eq:z*2} and \eqref{eq::notation1} respectively, there exists at least one coexisting equilibrium $(z^{'*1},z^{'*2})\gg \mathbf{0}$ in $\mathbf{E}$. Moreover, $z^{'*1}+z^{'*2}\leq \mathbf{1}$.
% \end{thm}

\begin{prop} \label{thm:coexist}
Consider \eqref{eq::sys_sis_dt_zmul} with $l=2$, $m=1$ and satisfying Assumptions \ref{ass:multi}-\ref{ass:h1m}. If $s_1(I-hD_f^k + hB_f^k) > 1$ for $k=1,2$; and $s_1(-hD_f^1 +h(I-Z^{*2}) B_f^1) > 0$ and $s_1(-hD_f^2 +h(I-Z^{*1}) B_f^2) > 0$ (notice the crossed-relation between viruses) with $z^{*1}$, $z^{*2}$ and $Z^{*1}$, $Z^{*2}$ defined as in \eqref{eq:z*1}, \eqref{eq:z*2} and \eqref{eq::notation1} respectively, then there exists at least one coexisting equilibrium $(z^{'*1},z^{'*2})\gg \mathbf{0}$ in $\mathcal{E}$. Moreover, $z^{'*1}+z^{'*2}\leq \mathbf{1}$.
\end{prop}
%\hjk{This is just a matter of taste, but for me the above notation has too many superscripts}

\begin{pf}
See Appendix \ref{app:17}.
\end{pf}

Next, we provide further detail for the case of two competing viruses. Sometimes, even though the reproduction numbers of both viruses are greater than 1, the dominant endemic equilibrium is still locally stable.

\begin{prop} \label{thm:win}
Under Assumption \ref{ass:multi}-\ref{ass:h1m}, with $l=2$ and $m=1$, for the model \eqref{eq::sys_sis_dt_zmul}, if $s_1(I-hD_f^k + hB_f^k) > 1$ for $k=1,2$ and $(D_f^1)^{-1}B_f^1 >(D_f^2)^{-1}B_f^2$, there exists exactly three equilibra: the healthy state, which is unstable; the dominant endemic equilibrium $(\mathbf{0},\hat{z}^{*2})$, with $\hat{z}^{*2} \gg \mathbf{0}$, which is unstable; the dominant endemic equilibrium $(\hat{z}^{*1},\mathbf{0})$, which is locally exponentially stable if $h< \frac{2}{\rho(J)}$, where $\hat{z}^{*1}>\hat{z}^{*2}$, 
\begin{equation}\label{eq:J}
    J=\left[ 
        \begin{matrix}
	(I-\hat{Z}^{*1})B^1_{f}-D^1_{f}-T & -T  \\
	\mathbf{0} & (I-\hat{Z}^{*1})B^2_{f}-D^2_{f}  \\
	\end{matrix}\right],
\end{equation}
and 
	\begin{equation*}
	    \begin{split}
        T=\left[ 
        \begin{matrix}
	\Dg(B^1_{f}x^{*1}) & \mathbf{0}  \\
	\mathbf{0} & \mathbf{0}  \\
	\end{matrix}
        \right]
        \end{split}
	\end{equation*}
\end{prop}

\begin{pf}
See Appendix \ref{app:18}.
\end{pf}

\subsection{Time-varying parameters}

Now, we consider the time-varying system \eqref{eq::sys_sis_dt_zmult}. Firstly, we investigate the healthy state behaviour of the time-varying model.

\begin{thm} \label{thm:mul11}
Consider the system \eqref{eq::sys_sis_dt_zmult} and suppose that there exists constant matrices $B^k_{f\max}$ and $D^k_{f\min}$ such that $0<B^k_f(t)<B^k_{f\max}$ and $D^k_{f\min}<D^k_f(t)\leq \mathbf{1}$ hold for every $k\in [l]$. Moreover, suppose that Assumptions \ref{ass:multi}-\ref{ass:bfm} hold for any $B^k_{f}(t)$ and $D^k_{f}(t)$ and $B^k_{f\max}$, in addition $D^k_{f\min}$ $s_1(I-hD_{f\min}^k + hB_{f\max}^k
) \leq 1$ holds for all $k\in [l]$, then the healthy
state is the unique equilibrium of system \eqref{eq::sys_sis_dt_zmul}, which is asymptotically stable with domain of attraction $\mathcal{E}$.
\end{thm}

\begin{pf}
See Appendix \ref{app:19}.
\end{pf}

Akin to the time invariant case, the time-varying model also has a dominant endemic equilibrium.

\begin{thm} \label{thm:mul12}
Consider the system \eqref{eq::sys_sis_dt_zmult} and suppose that there exists constant matrices $B^a_{f\max}$ and $D^a_{f\min}$ such that $0<B^a_f(t)<B^a_{f\max}$, $D^a_{f\min}<D^a_f(t)\leq \mathbf{1}$ and $s_1(I-hD_{f\min}^a + hB_{f\max}^a
) \leq 1$ hold for every $a\neq k$. Moreover, suppose that for the $k$-th virus, if $B^k_f(t)$ is symmetric for all $t\geq 0$ and $(B^k_f(t),D^k_f(t))\in \Omega_{z^*}$, where $\Omega_{z^*}$ is the set that concludes all the pairs $(B_f,D_f)$ that refer to the same endemic equilibrium $z^*$ for the time invariant system \eqref{eq::sys_sis_dt_z} (notice that $s_1(I-hD_{f}^a(t) + hB_{f}^a(t)
) >1$ holds already). Under Assumptions \ref{ass:multi}-\ref{ass:h1m}, 
\begin{itemize}
    \item [$i)$] the healthy state is asymptotically stable with domain of attraction $\mathcal{H}=\{A(t)=[Z^1(t),\dots,Z^l(t)]^{\top}|\\ z^a(t)\in\mathcal{D^\text{a}}, z^k(t)=\mathbf{0},\quad  \text{for any $a\in [l]$ and $a\neq k$}\}$;
    
    \item [$ii)$] the unique dominant endemic equilibrium\\ $(\mathbf{0},\dots,z^{k*},\dots,\mathbf{0})$ with $z^{k*} \gg \mathbf{0}$ is asymptotically stable with domain of attraction $\mathcal{E}\setminus \mathcal{H}$.
\end{itemize}

\end{thm}

\begin{pf}
See Appendix \ref{app:20}.
\end{pf}

{Theorems \ref{thm:mul11} and \ref{thm:mul12} are an extension of Theorems \ref{thm:6} and \ref{thm:7} into the multi-virus case. No previous studies on networked multi-virus SIS or SIWS models seem to have dealt with this problem before.}

\section{Simulations} \label{sec:sim}
In this section, we use several numerical examples to illustrate our analytical results and further show some potential properties of the system.

In all simulated scenarios, we consider a virus or multiple viruses spreading over a human contact network of {15} population nodes and its pathogen diffuses over 2 resource nodes. Both networks are randomly generated. The human contact network as well as the whole equivalent graph are set to be strongly connected. Each resource node is interconnected with at least one node in the human contact network. The spreading parameters are randomly generated from $[0,1]$. The sampling period $h$ is picked as $h=0.01$. We check the parameters strictly so that the assumptions  \ref{ass:single}-\ref{ass:h1} are satisfied for single-virus case and \ref{ass:multi}-\ref{ass:h1m} for multi-virus case.

Let $\Bar{x}^k$ denote the average infection ratio of $k$-th virus among all population, i.e. $\Bar{x}^k=\frac{1}{n}\sum_{i=1}^{3}x_i^k.$ Let $\Bar{w}^k$ denote the average contamination ratio among all resources, i.e. $\Bar{w}^k=\frac{1}{n}\sum_{i=1}^{2}w_i^k.$ Let $\rho$ denote $\rho=\rho(I-hD^k_f+hB^k_f)=s_1(I-hD^k_f+hB^k_f)$ for simplicity. The superscript $k$ is omitted in the single-virus case.

\subsection{Simulations of the time-invariant single-virus model} \label{sec:simsig}
Firstly, we show the simulation results for the single-virus time invariant case. 
\begin{figure*}
  \begin{subfigure}[t]{0.24\linewidth}
    \centering
    \includegraphics[height=3.5cm]{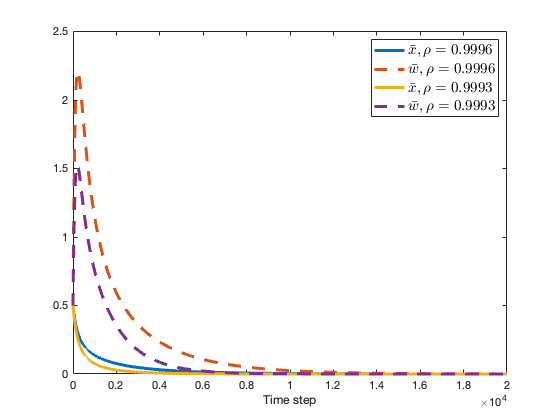}
    \captionsetup{width=.95\textwidth}
    \caption[width=3cm]{}
    \label{fig:sim_sig_healthy2}
  \end{subfigure}
  \begin{subfigure}[t]{0.24\linewidth}
    \centering
    \includegraphics[height=3.5cm]{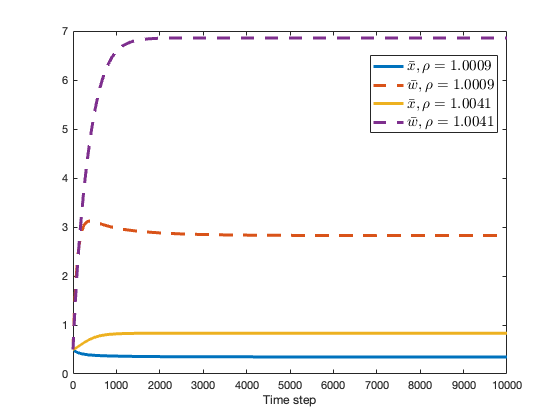}
    \captionsetup{width=.95\textwidth}
    \caption[width=3cm]{}
    \label{fig:sim_sig_end}
  \end{subfigure}
  \begin{subfigure}[t]{0.24\linewidth}
    \centering
\includegraphics[height=3.5cm]{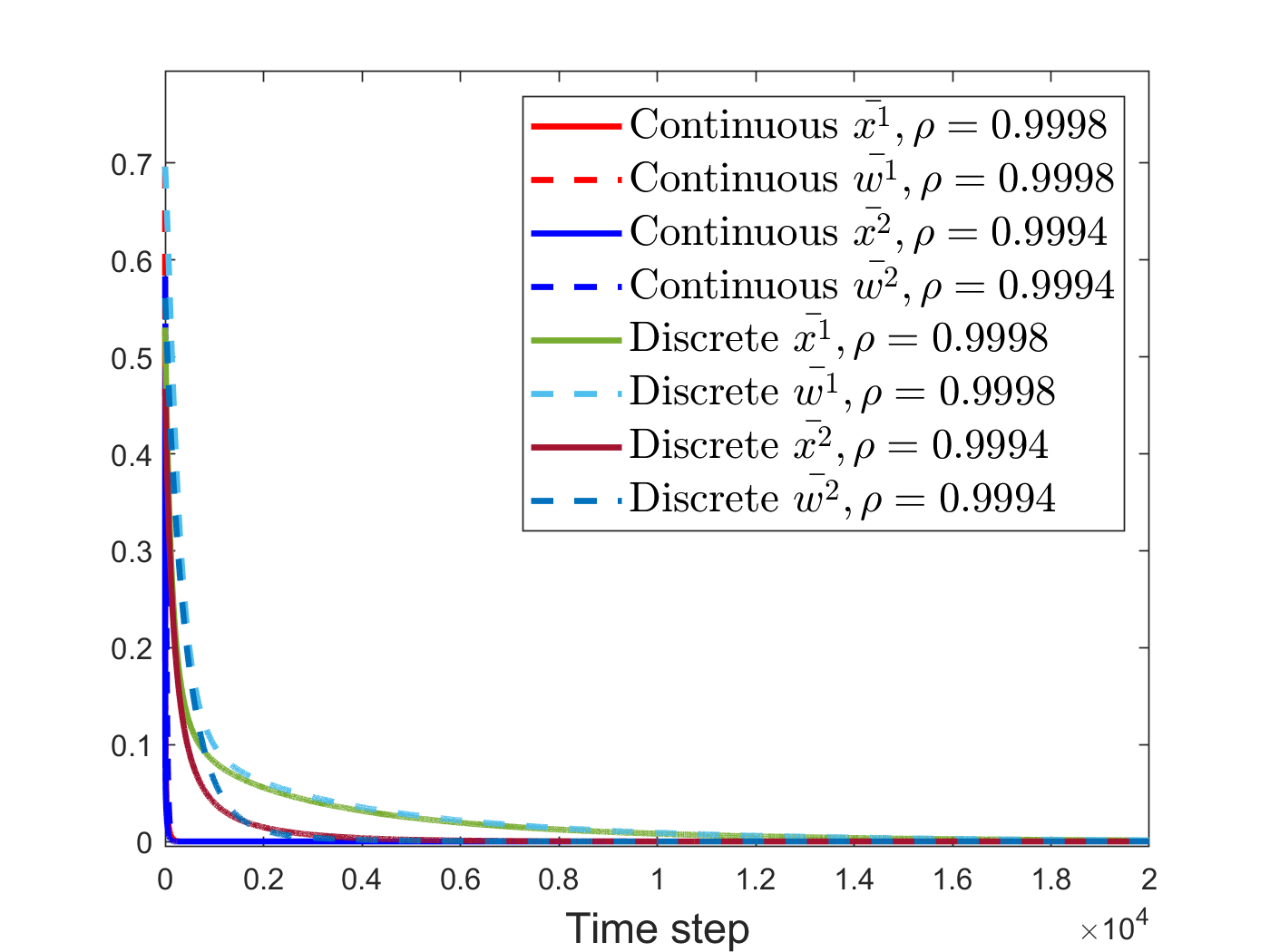}
\captionsetup{width=.95\textwidth}
    \caption[width=3cm]{}
    \label{fig:sim_mul_healthy}
  \end{subfigure}
  \begin{subfigure}[t]{0.24\linewidth}
    \centering
\includegraphics[height=3.5cm]{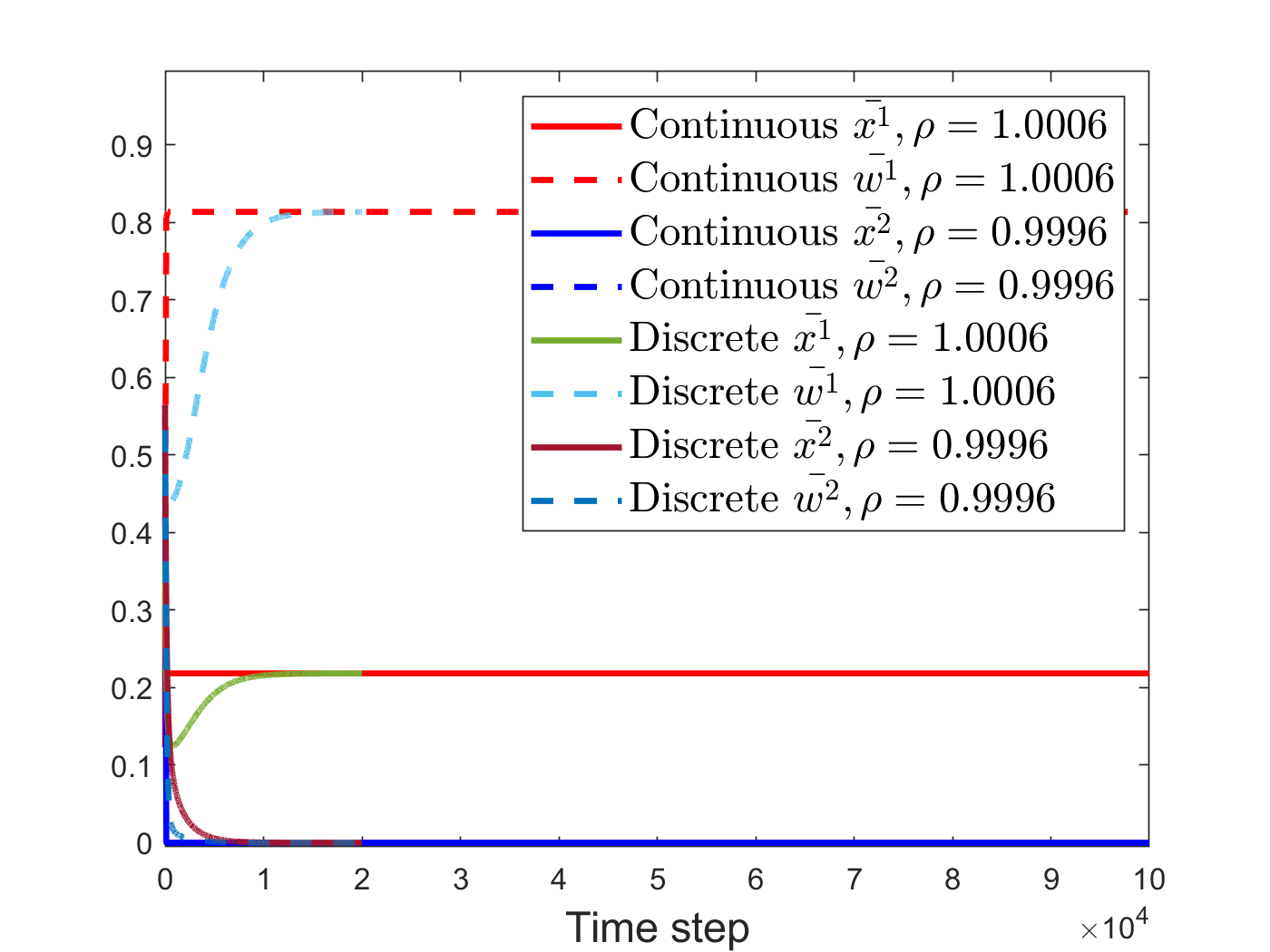}
\captionsetup{width=.95\textwidth}
\caption{}
    \label{fig:sim_mul_dom}
  \end{subfigure}
   \caption[width=3cm]{(a) Simulation results for the case $s_1(I-hD_f+hB_f){<}1$ from the same initial condition.
   (b) Simulation results for the case $s_1(I-hD_f+hB_f)>1.$
   (c) Simulation results for the case $s_1(I-hD^1_f+hB^1_f){<}1$ and $s_1(I-hD^2_f+hB^2_f){<}1.$
   (d) Simulation results for the case $s_1(I-hD^1_f+hB^1_f)>1$ and $s_1(I-hD^2_f+hB^2_f){<}1.$}
 \end{figure*}
 \begin{figure*}
  
  \begin{subfigure}[t]{0.24\linewidth}
    \centering
\includegraphics[height=3.5cm]{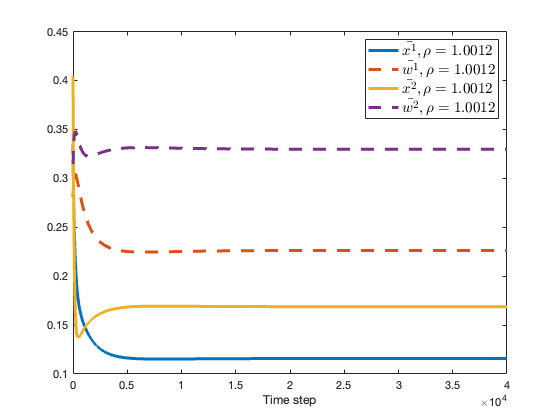}
\captionsetup{width=.95\textwidth}
    \caption{}
    \label{fig:sim_mul_coexist}
  \end{subfigure}
  \begin{subfigure}[t]{0.24\linewidth}
    \centering
\includegraphics[height=3.5cm]{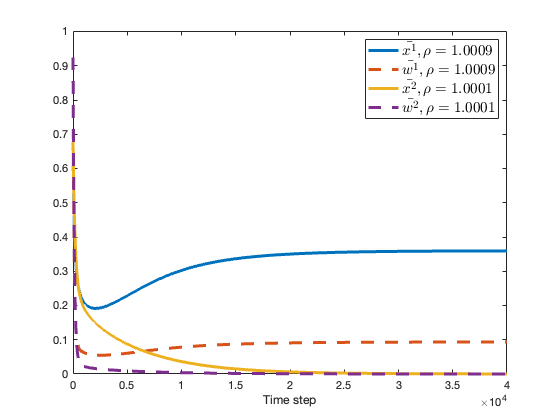}
\captionsetup{width=.95\textwidth}
    \caption{}
    \label{fig:sim_mul_com}
  \end{subfigure}
  \begin{subfigure}[t]{0.24\linewidth}
    \centering
\includegraphics[height=3.5cm]{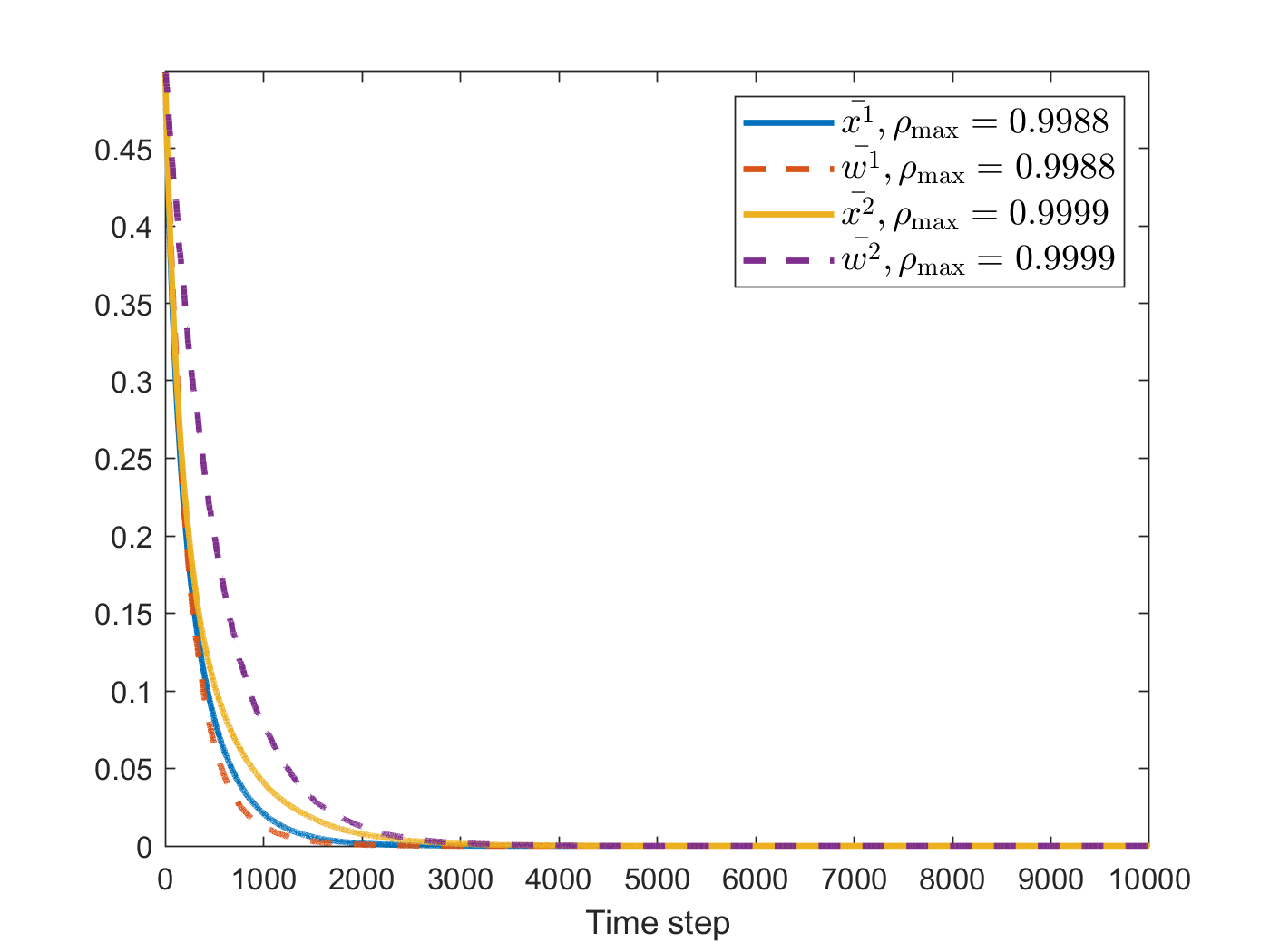}
\captionsetup{width=.95\textwidth}
    \caption{}
    \label{fig:sim_mul_ht}
\end{subfigure}
\begin{subfigure}[t]{0.24\linewidth}
    \centering
\includegraphics[height=3.5cm]{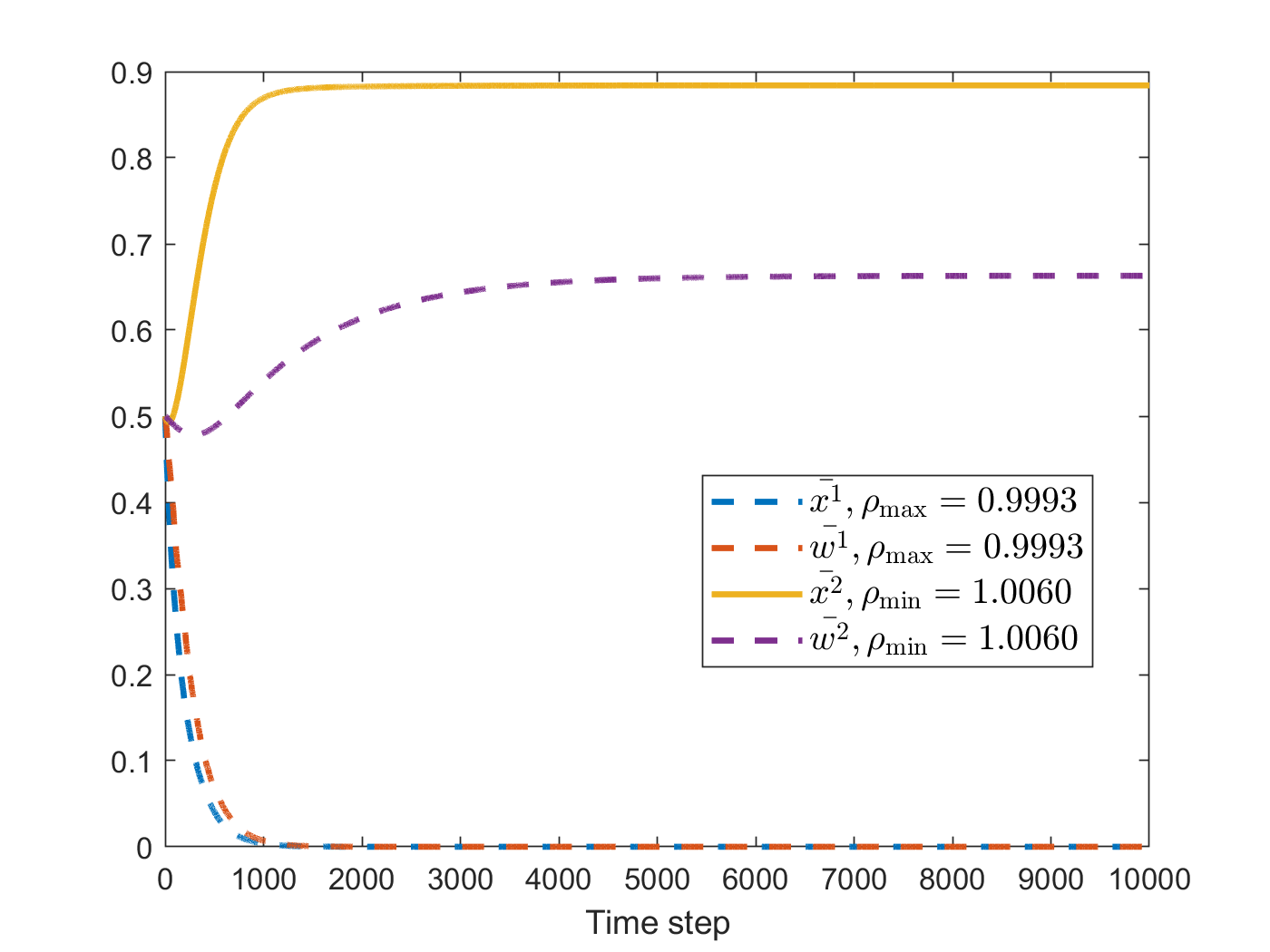}
\captionsetup{width=.95\textwidth}
    \caption{}
    \label{fig:sim_mul_endt}
\end{subfigure}
\caption{(a) Simulation results for the case $s_1(I-hD^1_f+hB^1_f) =s_1(I-hD^2_f+hB^2_f)>1.$
(b) Simulation results for the case $s_1(I-hD^1_f+hB^1_f)>1$ and $s_1(I-hD^2_f+hB^2_f)>1.$
(c) Simulation results for the case $s_1(I-hD_{f\min}^k + hB_{f\max}^k
) {<} 1$, for $k=1,2$.
(d) Simulation results for the case $s_1(I-hD_{f\min}^1 + hB_{f\max}^1
) {<} 1$ and $s_1(I-hD^2_f+hB^2_f)>1.$}
\end{figure*}
Figures \ref{fig:sim_sig_healthy2} show the simulation results for healthy state behaviour. We set all the initial infection level and contamination level to be $0.5$. Since $s_1(I-hD_f+hB_f)\leq1$, the system finally converges to its origin, which is in line with Theorem \ref{thm:hss}. Moreover, we can also observe that the greater $s_1(I-hD_f+hB_f)$ is, the slower the model converges to the origin. This conjecture is locally true because of Proposition \ref{prop:crate}. One of the future work is to validate whether this conjecture based on the observation holds true globally. Figure \ref{fig:sim_sig_end} shows the simulation results for the endemic behaviour with randomly generated initial condition. It is clear from Lemma \ref{lemma:endeqlemma} that the endemic equilibrium only depends on the spreading parameters and not related to the initial condition. Since $s_1(I-hD_f+hB_f)>1$, the system finally converges to its endemic equilibrium, which is in line with Theorem \ref{thm:end}. Moreover, we can also see that the greater $s_1(I-hD_f+hB_f)$ is, the greater contamination and infection level of the corresponding endemic equilibrium are. As above, whether this conjecture based on the observation holds true generally is not clear and needs future research.

\subsection{Simulations of the time invariant multi-virus model}\label{sec:sim1}
Next, we show the simulation results for the time invariant multi-virus case.
%\begin{figure}
%    \centering
%\includegraphics[height=6.5cm]{fig/healthymul.png}
%    \caption{Simulation results for the case $s_1(I-hD^1_f+hB^1_f)\leq1$ and $s_1(I-hD^2_f+hB^2_f)\leq1.$}
%    \label{fig:sim_mul_healthy}
%\end{figure}
%\begin{figure}
%   \centering
%\includegraphics[height=6.5cm]{fig/dominance.png}
%    \caption{Simulation results for the case $s_1(I-hD^1_f+hB^1_f)>1$ and $s_1(I-hD^2_f+hB^2_f)\leq1.$}
 %   \label{fig:sim_mul_dom}
%\end{figure}

We set the initial infection level and contamination level to be $0.5$. Figure \ref{fig:sim_mul_healthy} shows that the model converges to its origin when $s_1(I-hD^1_f+hB^1_f)<1$ and $s_1(I-hD^2_f+hB^2_f)<1,$ which is in line with Theorem \ref{thm:hmul}. We can observe that the virus with higher reproduction number will die out more slowly. Whether it is generally true remains an open question. Then, we use randomly generated initial condition. Figure \ref{fig:sim_mul_dom} shows that the model converges to its dominant endemic equilibrium when $s_1(I-hD^1_f+hB^1_f)>1$ and $s_1(I-hD^2_f+hB^2_f)\leq1,$ which is in line with Theorem \ref{thm:mul1}.
Figure \ref{fig:sim_mul_coexist} shows that the model converges to coexisting equilibrium when $s_1(I-hD^1_f+hB^1_f)=s_1(I-hD^2_f+hB^2_f)>1$. The condition of Theorem \ref{thm:coexist} is satisfied in this case. The simulation supports the existence of the coexisting equilibrium and suggests that it might be unique and asymptotically stable. One of the future work is to validate whether this conjecture based on the observation is true.
Figure \ref{fig:sim_mul_com} shows that the model converges to its dominant endemic equilibrium when $s_1(I-hD^1_f+hB^1_f)>1$, $s_1(I-hD^2_f+hB^2_f)>1$ and condition of Theorem \ref{thm:coexist} is not satisfied, which is in line with Theorem \ref{thm:win}. We also compare the results of our discrete-time system with the continuous-time counterpart. One can see from the figures \ref{fig:sim_mul_healthy} and \ref{fig:sim_mul_dom} that both solutions of continuous-time and discrete-time systems converge to the exact same equilibrium.

\subsection{Simulations of the time-varying multi-virus system}
In this subsection, we provide simulations for the time-varying system \eqref{eq::sys_sis_dt_zmult}. {Due to computation cost, we make a simulation now on a graph of 5 population nodes and 2 resource nodes.} We now allow the system parameters switched randomly among two parameters pair $(D_{f1},B_{f1})$ and $(D_{f2},B_{f2})$. We set the initial infection level and contamination level to be $0.5$ to better observe the convergence rate. {From further numerical studies, we still confirm that the model converges to the same equilibrium from a random initial value, but the figure is omitted here.}

\begin{figure}[t]
    \centering
\includegraphics[height=4.5cm]{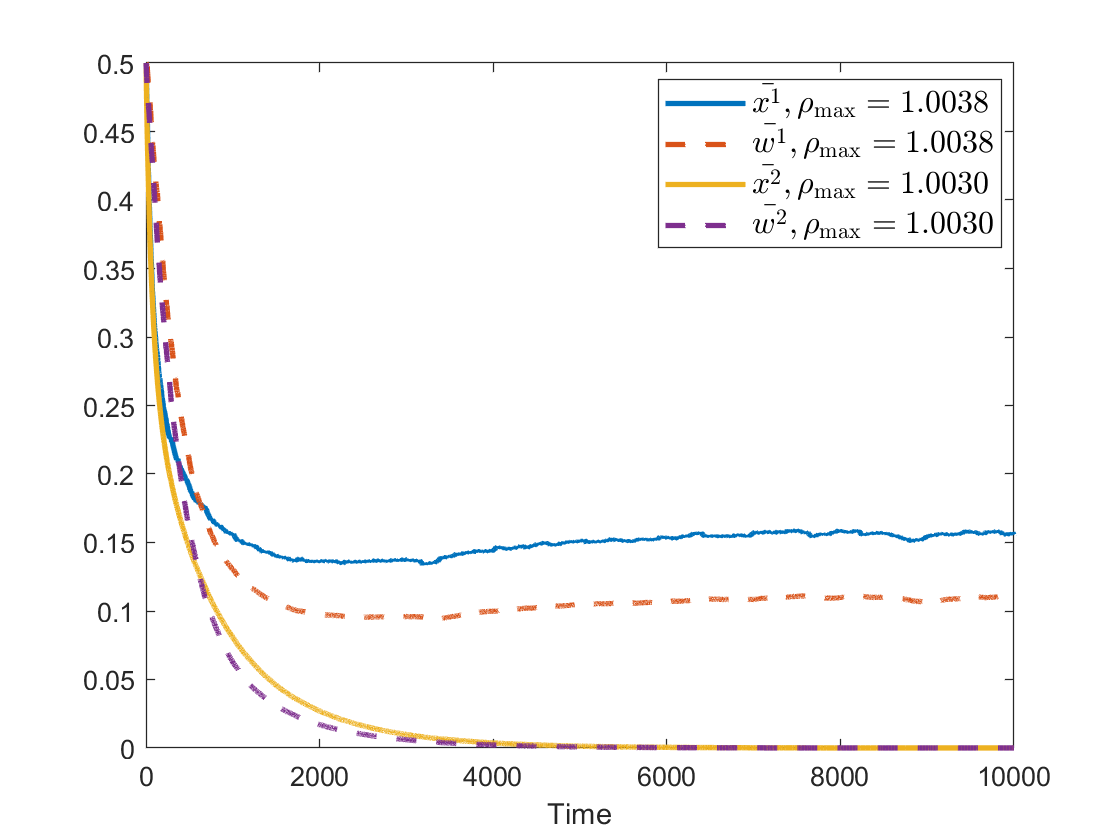}
\captionsetup{width=.95\textwidth}
    \caption{Simulation results for the case $s_1(I-hD^k_f+hB^k_f)<1$ but $s_1(I-hD_{f\min}^k + hB_{f\max}^k) > 1$, for $k=1,2.$}
    \label{fig:sim_mul_hts}
\end{figure}

Firstly, we set $s_1(I-hD_{f\min}^k + hB_{f\max}^k
) \leq 1$. 
Figure \ref{fig:sim_mul_ht} shows that if $s_1(I-hD_{f\min}^k + hB_{f\max}^k) \leq 1$ holds, the model converges to the healthy state, which is in line with the result of Theorem \ref{thm:mul11}. Similarly, we can observe that the virus with higher maximum production number will die out more slowly. 
{Figure \ref{fig:sim_mul_hts} illustrates that even if  $s_1(I-hD^1_f+hB^1_f)<1$ and $s_1(I-hD^2_f+hB^2_f)<1$ hold, once $s_1(I-hD_{f\min}^k + hB_{f\max}^k) >1$, the model does not converge to the healthy state. Instead, one virus dies out and the other oscillates around some value. This shows that the condition of Theorem \ref{thm:mul11} is general. Once violated, one observes another outcome. In future works we will study the system behaviour under such a condition.}

Then, we make a simulation on the endemic behaviour. We set the first virus remaining the setting same to simulation of Figure \ref{fig:sim_mul_ht} and set the spreading parameters of second virus homogeneous and changing over time proportionally so that the endemic equilibrium $x^*=1-\frac{\delta}{n\beta(1+m\hat{c})}$ and $w^*=n\hat{c}x^*$ will be fixed despite the time-varying parameters. Figure \ref{fig:sim_mul_endt} represents that once all the conditions in Theorem \ref{thm:mul12} are satisfied, the model converges to a dominant endemic equilibrium.
%\hjk{Comments for all figures: legends should be larger. Why are there different line widths? Please print one page and see that it looks OK. Maybe think of making the font of the captions smaller. It would be better to have time, and not timestep in the horizontal axis. One must specify that the vertical axis is showing two different quantities, one is a probability, or a fraction, while the other is a concentration}

%\hjk{There are 13 errors at the time I have written this, please try to take care of them.}

\section{Conclusion and Discussion}\label{sec:con}

In this paper, we have proposed a discrete-time networked layered SIWS model with multiple resources under both single virus and multiple viruses set-up. We have further generalized such models by considering time-varying spreading parameters. We have carried-out a full system's analysis on all proposed models, where we concentrated in the analysis of the healthy state behaviour and of the endemic behaviour. {We have solved several open questions regarding the multi-virus endemic behaviour of a discrete-time epidemic model and the existence of equilibria and their stability in the time-varying case.} Finally, we provided some numerical examples to illustrate our analytical results.

One of the future works will be to study the conjecture and the open questions mentioned in the previous section. Another potential future research direction should be validating the discrete-time model with real data and developing control strategies based on our models.

\bibliographystyle{plain}
\bibliography{bib}

\section{Appendix}
\subsection{Properties of a non-negative matrix}\label{app:prononneg}

We now recall the Perron-Frobenius Theorem for irreducible nonnegative matrices, which is one of the most important properties of irreducible nonnegative matrices.
\begin{lemma}{\cite[Theorem 2.7]{varga_springer_2000}} \label{lemma::perron_fr}
%Given that a square matrix $M$ is an irreducible nonnegative matrix, the following statements hold
 The following hold for a square irreducible nonnegative matrix $M$:
 \begin{itemize}
 	\item [$i)$] $M$ has a positive real eigenvalue equal to its spectral radius $\rho(M)$.
 	\item [$ii)$] $\rho(M)$ is a simple eigenvalue of $M$.
 	\item [$iii)$] There exist a unique right eigenvector $x \gg \mathbf{0}$ and a unique left eigenvector $y^{\top} \gg \mathbf{0}$ corresponding to $\rho(M)$.
% 	\item [$iv)$] $\rho(M)$ increases when any entry of $M$ increases.
% 	\item [$v)$] If $Q$ is an irreducible nonnegative matrix and $M > Q$, then $\rho(M) > \rho(Q)$.  
 \end{itemize} 
\end{lemma}
Other than the Perron-Frobenius Theorem, we recall the following lemmas showing the properties of nonnegative matrices which are necessary for this paper.
\begin{lemma}{\cite[Corollaries 8.1.29 and 8.1.30]{horn_cam_2013}} \label{lemma::nonneg_compare}
 Given a nonnegative matrix $M \in \mathbb{R}^{n \times n}$, a positive vector $x \in \mathbb{R}^{n}$ ($x \gg \mathbf{0}$) and a nonnegative scalar $\mu$, the following hold: 
 \begin{itemize}
     \item[$i)$] if $\mu x \ll Mx$, then $\mu < \rho(M)$,
     \item[$ii)$] if $\mu x \gg Mx$, then $\mu > \rho(M)$,
     \item[$iii)$] if $\mu x = Mx$, then $\mu = \rho(M)$.
 \end{itemize}
\end{lemma}

\begin{lemma} {\cite{rantzer2011distributed}} \label{lem:greater1}
 Suppose that $M$ is an irreducible non-negative matrix such that $s_1(M) < 1$. Then, there exists a positive diagonal matrix $P$ such that $M^\top PM-P$ is negative definite.
 \end{lemma}
 \begin{lemma}{\cite{pare2018analysis}} \label{lem:equal1}
 Suppose that $M$ is an irreducible non-negative matrix such that $s_1(M) = 1$. Then, there exists a positive diagonal matrix $P$ such that $M^\top PM-P$ is negative semi-definite.
 \end{lemma}
 
 \begin{lemma}{\cite[Proposition 1]{liu2019analysis}} \label{lemma:trans}
Suppose that $\Lambda$ is a negative diagonal matrix
in $\mathbb{R}_{n\times n}$ and $N$ is an irreducible nonnegative matrix in $\mathbb{R}_{n\times n}$ .
Let $M = \Lambda+N$. Then, $s_1(M) < 0$ if and only if $\rho(-\Lambda^{-1}N) <
1$, $s(M) = 0$ if and only if $\rho(-\Lambda^{-1}N) =
1$, and $s(M) > 0$
if and only if $\rho(-\Lambda^{-1}N) >
1$.
\end{lemma}

\begin{lemma}{\cite{varga1962iterative}, Lemma 2.6} \label{lemma:comp}
Suppose that $N$ is an irreducible
nonnegative matrix. If $M$ is a principal square submatrix of
$N$, then $\rho(M) < \rho(N)$.
\end{lemma}

\subsection{Proof of Lemma \ref{lemma:bound}} \label{app:1}

We can rewrite equation (\ref{eq::sys_sis_dt_x}) as 
\begin{align*} 
 x_i(t+1)&= x_i(t)(1-h\delta_{i})+(1-x_i(t))h\left[  \sum_{j=1}^{n} \beta_{ij} x_j(t)\right.\\
 &\left.+\sum_{j=1}^{m} \beta_{ij}^{w} w_j(t)\right].\\
\end{align*}
So, $x_i(t+1)$ is the convex combination of $(1-h\delta_{i})$ and $h  (\sum_{j=1}^{n} \beta_{ij} x_j(t)+\sum_{j=1}^{n} \beta_{ij}^{w} w_j(t))$, where the coefficients are strictly in the interval $[0,1]$. For $t=0$ and considering  Assumptions \ref{ass:single}, one obtains that $x_{i}(1) \in [0,1]$ holds. Next, assuming that $x_{i}(t) \in [0,1]$ holds, it follows that $x_{i}(t+1) \in [0,1]$ also holds. Thus, by induction, we prove that $x_{i}(t) \in [0,1]$ for all $t\geq 0$.

Similarly, we can also rewrite equation \eqref{eq::sys_sis_dt_w} as
\begin{equation*} 
 w_j(t+1)= w_j(t)(1-h\delta_{j}^{w})+h\frac{\delta_{j}^{w}}{\delta_{j}^{w}}\sum_{k=1}^{n}c_{jk}^{w}x_{k}(t),
\end{equation*}
where $w_j(t+1)$ is the convex combination of $w_j(t)$ and $\frac{1}{\delta_{j}^{w}}\sum_{k=1}^{n}c_{jk}^{w}x_{k}(t),$ with the coefficients being strictly in the set $[0,1]$. By an analogous induction argument as above, and using Assumption \ref{ass:single}, it follows that 
$w_{j}(t) \in [0,w_{\max}]$ for all $j \in [m]$ holds, for all $t\geq 0$.

\subsection{Proof of Lemma \ref{lem:boundt}}\label{app:2}

% \begin{pf}
% Firstly, we have
% \begin{equation*} 
% \begin{split}
%     x_i(t+1)&=x_i(t)+h[ (1-x_i(t)) (\sum_{j=1}^{n} \beta_{ij}(t) x_j(t)\\
%  &+\sum_{j=1}^{m} \beta_{ij}^{w}(t) w_j(t)) -\delta_i(t) x_i(t) ]\\
%  &\leq x_i(t)(1-h\delta_{i}_{\min})\\
%  &+(1-x_i(t))h\left[  \sum_{j=1}^{n} \beta_{ij,\max} x_j(t)\right.\\
%  &\left.+\sum_{j=1}^{m} \beta_{ij,\max}^{w} w_j(t) \right].
% \end{split}
% \end{equation*}

% Following the same proof of Lemma \ref{lemma:bound}, we have $x(t) \in [0,1]$ for all $t \geq 0$.

% Now, consider the dynamic of $w(t)$, 
% \begin{equation*} 
% \begin{split}
%      w_j(t+1)&=w_j(t)+h\left(-\delta_{j}^{w}(t)w_{j}(t)+\sum_{k=1}^{n}c_{jk}^{w}(t)x_{k}(t)\right)\\
%      &\leq w_j(t)(1-h\delta_{j\min}^{w})+h\frac{\delta_{j\min}^{w}}{\delta_{j\min}^{w}}\sum_{k=1}^{n}c_{jk\max}^{w}x_{k}(t).
% \end{split}
% \end{equation*}
% Following the same proof of Lemma \ref{lemma:bound}, we have $w(t) \in [0,w_{\max}]$ for all $t \geq 0$.
% \end{pf}

 It suffices to notice that due to the hypothesis of the Lemma we have:
\begin{equation*} 
\begin{split}
    x_i(t+1)&=x_i(t)+h\left[ (1-x_i(t)) \left(\sum_{j=1}^{n} \beta_{ij}(t) x_j(t)\right.\right.\\
 &\left.\left.+\sum_{j=1}^{m} \beta_{ij}^{w}(t) w_j(t)\right) -\delta_i(t) x_i(t) \right]\\
 &\leq x_i(t)(1-h\delta_{i\min})\\
 &+(1-x_i(t))h\left[  \sum_{j=1}^{n} \beta_{ij\max} x_j(t)\right.\\
 &\left.+\sum_{j=1}^{m} \beta_{ij\max}^{w} w_j(t) \right],
\end{split}
\end{equation*}
and
\begin{equation*} 
\begin{split}
     w_j(t+1)&=w_j(t)+h\left(-\delta_{j}^{w}(t)w_{j}(t)+\sum_{k=1}^{n}c_{jk}^{w}(t)x_{k}(t)\right)\\
     &\leq w_j(t)(1-h\delta_{j\min}^{w})+h\frac{\delta_{j\min}^{w}}{\delta_{j\min}^{w}}\sum_{k=1}^{n}c_{jk\max}^{w}x_{k}(t),
\end{split}
\end{equation*}
where $\beta_{ij\max}$,  $\beta_{ij\max}^w$, and $c_{jk\max}^w$ are the corresponding entries of $B_{f\max}$ while $\delta_{i\min}$ and $\delta_{i\min}^w$ of $D_{f\min}$. Then, one can follow analogous steps as in the proof of Lemma \ref{lemma:bound} to obtain the result.

\subsection{Proof of Theorem \ref{thm:hss}} \label{app:3}

 Firstly, we show the statement i) is true.
 Notice that the system equation \eqref{eq::sys_sis_dt_z} in matrix form is basically the same as \cite[Equation (7)]{pare2018analysis}, where $B_f$ and $D_f$ are somewhat extended and contain the information of each resource. Observe that, due to our assumptions, $D_f$ is still a positive diagonal matrix and $B_{f}$ is still a non-negative irreducible matrix. Following a similar proof as for \cite[Theorem 1]{pare2018analysis}, we can obtain the result. We now sketch some of the main arguments and highlight the differences.

 We consider the case $s_{1}(I-hD_f + hB_f) < 1$ and let $M=I-hD_f+hB_f$.
 By Lemma \ref{lem:greater1}, there exists a $P_1$ such that $M^\top P_1 M-P_1$ is negative definite We define the Lyapunov function $V_1(z(t)) = (z(t))^\top P_1 (z(t))$. Let $\Bar{M}=I+h((I-Z(t))B_f-D_f)$. Thus,
 \begin{equation}\label{eq::lyapunov}
 \begin{split}
     \Delta V_1(z(t+1))&=V_1(z(t+1))-V_1(z(t))\\
     &=(z(t))^\top \Bar{M}^{\top} P_1  \Bar{M} (z(t))-(z(t))^\top P_1 (z(t))\\
     &=(z(t))^\top ({M}^{\top} P_1  {M}-P_1) (z(t))\\
     &- 2h(z(t))^\top {B_f}^{\top} Z(t) P_1  {M} (z(t))\\
     &+h^2(z(t))^\top {B_f}^{\top} Z(t) P_1  Z(t) B_f (z(t))\\
     &\leq -2h(z(t))^\top {B_f}^{\top} Z(t) P_1  {M} (z(t))\\
     &+h^2(z(t))^\top {B_f}^{\top} Z(t) P_1  Z(t) B_f (z(t)).
 \end{split}
 \end{equation}
The last inequality holds as equality if and only if $z(t)=0$, since ${M}^{\top} P_1  {M}-P_1$ is negative definite.
 
 Following the same calculation as in \cite{pare2018analysis}, we have $\Delta V_1(z(t+1))\leq -2h(z(t))^\top {B_f}^{\top} Z(t) P_1  {M} (z(t))+h^2(z(t))^\top {B_f}^{\top} Z(t) P_1  Z(t) B_f (z(t)) \leq 0$. Besides, $\Delta V_1(z(t+1))<0$ for all $z(t)\neq 0$. This ensures that the model converges asymptotically to the healthy state under this circumstance.

Next, we consider the case $s_{1}(I-hD_f + hB_f) = 1$. By Lemma \ref{lem:equal1}, there exists a $P_2$ such that $M^\top P_2 M-P_2$ is negative semi-definite. We define the Lyapunov function $V_2(z(t)) = (z(t))^\top P_2 (z(t))$.

We can observe that calculation \eqref{eq::lyapunov} still holds in this case but only requires $x(t)=\mathbf{0}$, because $M^\top P_2 M-P_2$ is now negative semi-definite. Following the same calculation as in \cite{pare2018analysis}, we have $\Delta V_2(z(t+1))\leq -2h(z(t))^\top {B_f}^{\top} Z(t) P_2  {M} (z(t))+h^2(z(t))^\top {B_f}^{\top} Z(t) P_2  Z(t) B_f (z(t))\\ \leq -h(z(t))^\top {B_f}^{\top} Z(t) P_2  {M} (z(t))  \leq 0$. Besides, notice that $Z(t)=\left[ 
        \begin{matrix}
	\Dg(x(t)) & \mathbf{0}  \\
	\mathbf{0} & \mathbf{0}  \\
	\end{matrix}
        \right]$ and accordingly $\Delta V_2(z(t+1))\leq 0$ holds as an equality if and only if $x(t)= \mathbf{0}$. According to the Lasalle's principle, the system converges asymptotically to the largest invariant set, where $\Delta V_2(z(t+1))=0$. This set is in the form of  $\{z(t)=[x(t)^\top,w(t)^\top]^{\top}\,|\,x(t)=\mathbf{0}, \,  w(t)\in [0,w_{\max}]^{m}\}$. This ensures that  $x(t)$ converges to zero. Recall that $w(t+1)= w(t)+h\left(  -D_{w}w(t)+C_{w}x(t)   \right)$. If $x(t)$ converges to zero, $w(t)$ is governed by a stable linear system plus a vanishing part of $x(t)$. This implies that $w(t)$ also converges to zero.
        
        Secondly, we prove that the statement ii) is true. We can follow the same proof of Theorem 2 in \cite{pare2021multi}, since $D_f$ is still a positive diagonal matrix and $B_{f}$ is still a non-negative irreducible matrix. Furthermore, the matrix $A_w$ is regarded as zero matrix for our case and assumption 2 of \cite{pare2021multi} is also satisfied. For the proof of a single resource case, interested readers can also refer to the proof of Theorem 3 in \cite{janson2020networked}. The main idea of the proof of existence of the endemic equilibrium is firstly to construct an auxiliary continuous map and apply the Brouwer’s fixed-point theorem. The uniqueness of the endemic equilibrium is proved by way of contradiction.
        
        Finally, the statement iii) is the direct consequence of the statements i) and ii).

\subsection{Proof of Proposition \ref{prop:2}}
\label{app:5}

We will need the following lemma for the proof.
\begin{lemma} \label{lemma:endeqlemma}
The endemic equilibrium $z^*=[x^*,w^*]^{\top} \gg 0$ of system \eqref{eq::sys_sis_dt_z}, if exists, must satisfy
\begin{equation}\label{eq:end1}
    \frac{\delta_{i}x_{i}^*}{1-x_{i}^*}=\sum_{j=1}^{n}\left(\sum_{k=1}^{m}(\beta^w_{ij}\hat{c}_{kj})+\beta_{ij}\right)x_{j}^*,
\end{equation}

and
\begin{equation}\label{eq:end2}
    w^*_{j}=\sum_{k=1}^{n}\hat{c}_{jk}x_{k}^*, 
\end{equation}
where $\hat{c}_{jk}^w\coloneqq c_{jk}^w/\delta_{j}^w$.
\end{lemma}

\begin{pf}
By the definition of an equilibrium, we have $z(t+1)=z(t)=z^*$. So, according to \eqref{eq::sys_sis_dt_w}, we have: 
\begin{equation*} 
 h\left(-\delta_{j}^{w}w_{j}^*+\sum_{k=1}^{n}c_{jk}^{w}x_{k}^*\right)=0.
\end{equation*}
Then, 
\begin{equation*}
    w^*_{j}=\sum_{k=1}^{n}\hat{c}_{jk}x_{k}^*.
\end{equation*}

Similarly, according to \eqref{eq::sys_sis_dt_x}, we obtain: 
\begin{equation*}
    h\left[ (1-x_i^*) \left(\sum_{j=1}^{N} \beta_{ij} x_j^*+\sum_{j=1}^{N} \beta_{ij}^{w} w_j^*\right) -\delta_i x_i^* \right]=0.
\end{equation*}
Then, 
\begin{equation} \label{eq::lemma6}
    \frac{\delta_{i}x_{i}^*}{1-x_{i}^*}=\left(  \sum_{j=1}^{n} \beta_{ij} x_j^*+\sum_{j=1}^{m} \beta_{ij}^{w} w_j^*  \right).
\end{equation}
By using \eqref{eq:end2}, we prove that \eqref{eq:end1} holds. That completes the proof.
\end{pf}

Now, we can rewrite \eqref{eq::sys_sis_dt_x} as 
\begin{equation}\label{eq:p21}
    \begin{split}
         e_i(t+1) &= x_i(t)(1-h\delta_{i})+(1-x_i(t))h\left[  \sum_{j=1}^{n} \beta_{ij} x_j(t)\right.\\
& \left.+\sum_{j=1}^{m} \beta_{ij}^{w} w_j(t)\right]-x_{i}^*.
    \end{split}
\end{equation}

%Since we have equation \eqref{eq:end1}, we can reformulate it as
Next, notice that \eqref{eq:end1} can be reformulated as:
\begin{equation*}
   h\delta_{i}\frac{x_{i}^*}{1-x_{i}^*}=h\sum_{j=1}^{n}\left(\sum_{k=1}^{m}(\beta^w_{ij}\hat{c}_{kj})+\beta_{ij}\right)x_{j}^*;
\end{equation*}

\begin{equation*}
   h\delta_{i}x_{i}^*=h(1-x_{i}^*)\sum_{j=1}^{n}\left(\sum_{k=1}^{m}(\beta^w_{ij}\hat{c}_{kj})+\beta_{ij}\right)x_{j}^*;
\end{equation*}

\begin{equation}\label{eq:p22}
   x_{i}^*=(1-x_{i}^*)h\sum_{j=1}^{n}\left(\sum_{k=1}^{m}(\beta^w_{ij}\hat{c}_{kj})+\beta_{ij}\right)x_{j}^*+(1-h\delta_{i})x_{i}^*.
\end{equation}

%\hjk{probably the second equation of the above can be deleted}\csx{if you think this calculation is obvious, we can consider to delete it}

By plugging \eqref{eq:p22} into \eqref{eq:p21}, we have
\begin{equation*}
   \begin{split}
         e_i(t+1) &=
         %e_i(t)(1-h\delta_{i})+(1-x_i(t))h\left[  \sum_{j=1}^{n} \beta_{ij} x_j(t)\right.\\
%&\left.+\sum_{j=1}^{m} \beta_{ij}^{w} w_j(t) \right] \\
%&- (1-x_{i}^*)h\sum_{j=1}^{n}\left(\sum_{k=1}^{m}(\beta^w_{ij}\hat{c}_{kj})+\beta_{ij}\right)x_{j}^\\
 e_i(t)(1-h\delta_{i})+ h\left(\sum_{j=1}^{m} \beta_{ij}^{w} (w_{j}(t)-w^*_{j})\right)\\
&+h\sum_{j=1}^{n} \beta_{ij} e_j(t)-h\sum_{j=1}^{n} \beta_{ij}(x_{i}(t)x_{j}(t)-x_{i}^*x_{j}^*)\\
&-h\left(\sum_{j=1}^{m} \beta_{ij}^{w} (x_{i}(t)w_{j}(t)-x_{i}^*w^*_{j})\right)
    \end{split}
\end{equation*}
%\hjk{I think the first equation can be omitted as these are straightforward computations}

Then, we look at the term $x_{i}(t)x_{j}(t)-x_{i}^*x_{j}^*$. We can observe that $x_{i}(t)x_{j}(t)-x_{i}^*x_{j}^*=(x_{i}(t)-x_{i}^*)(x_{j}(t)-x_{j}^*)+(x_{i}(t)-x_{i}^*)x_{j}^*+x_{i}^*(x_{j}(t)-x_{j}^*)$. Similarly, $x_{i}(t)w_{j}(t)-x_{i}^*w_{j}^*=(x_{i}(t)-x_{i}^*)(w_{j}(t)-w_{j}^*)+(x_{i}(t)-x_{i}^*)w_{j}^*+x_{i}^*(w_{j}(t)-w_{j}^*)$. Then
\begin{equation*}
   \begin{split}
         e_i(t+1) 
%&= e_i(t)(1-h\delta_{i})+ h(\sum_{j=1}^{m} \beta_{ij}^{w} f_{j}(t))\\
%&+h\sum_{j=1}^{n} \beta_{ij} e_j(t)\\
%&-h\sum_{j=1}^{n} \beta_{ij}(e_{i}(t)e_{j}(t)+e_{i}(t)x_{j}^*+x_{i}^*e_{j}(t))\\
%&-h(\sum_{j=1}^{m} \beta_{ij}^{w} (e_{i}(t)f_{j}(t)+e_{i}(t)w_{j}^*+x_{i}^*f_{j}(t))\\
&=\left(1-h\delta_{i}-h\sum_{j=1}^{n} \beta_{ij} x_j^*-h\sum_{j=1}^{m} \beta_{ij}^{w} w_{j}^*\right)e_{i}(t)\\
&+ (1-x_{i}^*)\left(h\sum_{j=1}^{n} \beta_{ij} e_j(t)+h\sum_{j=1}^{m} \beta_{ij}^{w} f_{j}(t)\right)\\
&-e_{i}(t)h\sum_{j=1}^{n} \beta_{ij} e_j(t)-e_{i}(t)h\sum_{j=1}^{m} \beta_{ij}^{w} f_{j}(t).
    \end{split}
\end{equation*}
%\hjk{Similarly, I think the first equation can be omitted}

From \eqref{eq:end1}, we have 
\begin{equation*}
   h\delta_{i}(\frac{1}{1-x_{i}^*}-1)=h\sum_{k=1}^{m}\beta^w_{ik}w_{k}^*+h\sum_{j=1}^{n}\beta_{ij}x_{j}^*.
\end{equation*}

For the pathogen error dynamics, we have 
\begin{equation*}
\begin{split}
    f_{j}(t+1)&=f_{j}(t)-h\delta_{j}^w w_{j}(t)+h\delta_{j}^w\sum_{k=1}^{n}\hat{c}_{jk}^{w}w_{k}(t)\\
    &=(1-h\delta_{j}^w)f_j(t)-h\delta_{j}^w w_j^*+h\delta_{j}^w\sum_{k=1}^{n}\hat{c}_{jk}^{w}w_{k}(t)\\
    &= (1-h\delta_{j}^{w})f_{j}(t)+h\sum_{k=1}^{n}c_{jk}^{w}e_{k}(t).
\end{split}
\end{equation*}

Thus, we get 
\begin{align}
     &e_i(t+1) = \left(1+\frac{h\delta_{i}}{x_{i}^*-1}\right)e_i(t) \notag \\
     &+(1-x_{i}^*-e_{i}(t))h\left(\sum_{j=1}^{n}\beta_{ij}e_{j}(t)+\sum_{j=1}^{m}\beta_{ij}^{w}f_{j}(t)\right),\label{eq::sys_sis_dt_xe} \\
     &f_j(t+1)= (1-h\delta_{j}^{w})f_{j}(t)+h\sum_{k=1}^{n}c_{jk}^{w}e_{k}(t).\label{eq::sys_sis_dt_we}
\end{align}

Finally, we can simplify the system into the matrix form corresponding to \eqref{eq:errormatrix}.

\subsection{Proof of Theorem \ref{thm:end}} \label{app:7}

Firstly, we need the following Lemma.

\begin{lemma} \label{lem:11}
Let \begin{equation} \label{eq:f}
    F=\left[ 
        \begin{matrix}
	I-\Dg\left(\frac{h\delta_1}{1-x_i^*}\right)+hB & hB_w  \\
	hC_w & I-hD_{w}  \\
	\end{matrix}\right]. 
\end{equation}
Then the matrix $F$ is non-negative irreducible under Assumptions \ref{ass:single}-\ref{ass:h1}. 
\end{lemma}

\begin{pf}
Under the Assumptions, we have $hC_w$, $hB_w$ and $I-hD_{w}$ are all non-negative matrices. Since $B_{f}$ (recall \eqref{eq::notation}) is irreducible under Assumption \ref{ass:bf}, $F$ must also be irreducible.
Next, we take a look at the matrix
\begin{equation*}
    \left(I-\Dg\left(\frac{h\delta_i}{1-x_i^*}\right)+hB\right)_{ij}=\left\{ 
       \begin{matrix}
     1-\frac{h\delta_i}{1-x_i^*}+h\beta_{ii}, & i = j \\
         h \beta_{ij}, & i \neq j
       \end{matrix}
     \right.
\end{equation*}
We see that $\beta_{ij}\geq0$, so it remains to show that $1-\frac{h\delta_i}{1-x_i^*}+h\beta_{ii}\geq 0$. In fact
\begin{equation*}
    \begin{split}
        1-\frac{h\delta_i}{1-x_i^*}+h\beta_{ii}\geq 1-\frac{h\delta_i}{1-x_i^*},
    \end{split}
\end{equation*}
So we can equivalently show that $1-\frac{h\delta_i}{1-x_i^*}\geq 0$, which is further equivalent to $x_i^*\leq 1-h\delta_i$. By using \eqref{eq:end1}, we obtain
\begin{equation*}
    x_{i}^*=1-\frac{h\delta_{i}}{h\left(  \sum_{j=1}^{n} \beta_{ij} x_j^*+\sum_{j=1}^{m} \beta_{ij}^{w} w_j^*  \right)+h\delta_{i}}
\end{equation*}
So $x_{i}^*\leq 1-h\delta_i$ holds by Assumption \ref{ass:h1}. That finishes the proof.
\end{pf}

Let $\mu = [1,\frac{x^*_2}{x^*_1},\ldots,\frac{x^*_n}{x^*_1},\frac{w^*_1}{x^*_1},\dots, \frac{w^*_m}{x^*_1}]^{\top}$. Notice that
 \begin{equation}
  F \mu = \mu.
 \end{equation}

 Since $F$ is an irreducible nonnegative matrix and $\mu \gg \mathbf{0}$, it follows that $\rho(F) = 1$ by Lemma \ref{lemma::nonneg_compare}. According to Lemma \ref{lemma::perron_fr}, there exists a positive left eigenvector $v^{\top}$ such that $v^{\top} F = v^{\top}$. Then we can consider the following two cases.
 
 \emph{Case 1:} If $z(0) = \mathbf{0}$, we can directly obtain that the system stays at the healthy state and doesn't converge to the endemic equilibrium.  Indeed, the Jacobian matrix of \eqref{eq::sys_sis_dt_z} at the origin is $J(\mathbf{0},\mathbf{0})=\left[ 
        \begin{matrix}
	I+hB-hD & hB_{w}  \\
	hC_{w} & I-hD_w  \\
	\end{matrix}
        \right]=I+hB_f-hD_f.$ Since $s_{1}(I-hD_f + hB_f) > 1$, it follows that the healthy state is locally unstable.
Moreover, if we have $z(t+1) = \mathbf{0}$ at time $t+1$, we have
\begin{equation*} 
0= x_i(t)(1-h\delta_{i})+(1-x_i(t))h\left[  \sum_{j=1}^{n} \beta_{ij} x_j(t)
 +\sum_{j=1}^{m} \beta_{ij}^{w} w_j(t) \right]
\end{equation*}
and
\begin{equation*} 
 0= w_j(t)(1-h\delta_{j}^{w})+h\sum_{k=1}^{n}c_{jk}^{w}x_{k}(t).
\end{equation*}

Notice that all parameters in the equations are non-negative and there exists at least one $ \beta_{ij}^{w}>0$, $ \beta_{ij}>0$ or $c_{jk}^{w}>0$. It follows that $z(t) = \mathbf{0}$ is the only possible solution. Following the same proof, we can obtain $z(t-1) = \mathbf{0}$. We can always consider the previous time step. By induction, we obtain that $z(t) = \mathbf{0}$ if and only if $z(0) = \mathbf{0}$. Thus, $z(t)=\mathbf{0}, \forall t \geq 0$ if and only if $z(0)=\mathbf{0}$. 

\emph{Case 2:} If $z(0) \neq \mathbf{0}$, we show that there must exists a time $s$ such that 
 $x(s) \gg \mathbf{0}$. 
 
 %Since $\mathcal{G(B_{\text{f}})}$ as well as the human contact network must be strongly connected.
 Recall that $\mathcal{G(B_{\text{f}})}$ is strongly connected. %The human contact network is also strongly connected because resources are not connected directly with each other.
 Thus, there exists at least one $ \beta_{ij}^{w}>0$, $ \beta_{ij}>0$ or $c_{jk}^{w}>0$. By Assumption \ref{ass:single}, it holds that $1-h \delta_i \geq 0$ and $1-h \delta_i^w \geq 0$. Let $z_l(0) > 0$  for certain $l \in \mathcal{V(B_\text{f})}$. According to the discussion of case 1, we have $z_l(t) > 0$, for all $t \geq 0$.
Then, we consider the worst case. Without loss of generality, we assume only $z_p(0) > 0$, for a node $p \in \mathcal{V(B_{\text{f}})}$ and all other entries of $z(0)$ are $0$. Based on the conclusion of case 1, $z_p(t)$ stays positive for any $t \geq 0$. 
 Since $\mathcal{G(B_{\text{f}})}$ is strongly connected, we can always find a node $q$ such that there exists an edge from $q$ to $p$, i.e., $ \beta_{ij}^{w}>0$, $ \beta_{ij}>0$ or $c_{jk}^{w}>0$. It follows that $z_q(1)>0$. Similarly, all other entries will become positive in finite time steps due to the strong connectivity of the whole equivalent graph $\mathcal{G(B_{\text{f}})}$ that combines the human-contact network and resource network. 
 
 Next, let us consider the following auxiliary system for any $t \geq s$. 
 \begin{equation} \label{eq::sys_aux_y}
  y(t+1) = \Phi(t) y(t),
 \end{equation}
 with initial condition $y_i(s) = |e_i(s)|, \forall i \in \mathcal{V(B_{\text{f}})}$. We also have $\hat{e}(t+1)= \phi(t)\hat{e}(t)$. Let $\Phi(t)=\phi(x(t))$. We can observe that $\Phi(t)=\phi(x(t))$ is upper bounded by the matrix $F$. Since it is a nonnegative matrix, it follows that $-y(t) \leq \hat{e}(t)\leq y(t), \forall t \in \mathbb{N}$.
 
 The endemic equilibrium is asymptotically stable, if the origin of the system~\eqref{eq:errormatrix} is asymptotically stable. Thus, it remains to show that the origin of the system~\eqref{eq:errormatrix} is asymptotically stable. It is equivalent to show that the origin of the system~\eqref{eq::sys_aux_y} is asymptotically stable. %\hjk{don't you mean here (45)?}
 
 Define the Lyapunov function $V(t) = v^{\top}y(t)$, where $v^{\top}$ is the positive left eigenvector of $F$. We can calculate increment of the Lyapunov function as follows.
 \begin{equation}
  \begin{aligned}
   \Delta V(t) &= V(t+1)-V(t) = v^{\top} (\Phi(t)-I) y \\
               &= v^{\top}(\Phi(t)-F)y \\
               &= -h v^{\top} \left[ 
        \begin{matrix}
	\Dg(x(t)) & \Dg(x(t))  \\
	\mathbf{0} & \mathbf{0}  \\
	\end{matrix}\right]B_{f} y \\
               &\leq 0. 
  \end{aligned}
 \end{equation}
 Since $z,v \gg \mathbf{0}$ and the sum of each row in $B_f$ is positive, it follows that $\Delta V = 0$ if and only if $y_i =0$ for all $i\in [n]$. The model converges to the largest invariant set where $y_i =0$ for all $i\in [n]$. Since $y_j(t+1)= (1-h\delta_{j}^{w})y_{j}(t)+h\sum_{k=1}^{n}c_{jk}^{w}y_{k}(t)$ for all $n+1 \geq j \geq n+m$. $y_j(t+1)= (1-h\delta_{j}^{w})y_{j}(t)$ is a stable linear system and the term $h\sum_{k=1}^{n}c_{jk}^{w}y_{k}(t)$ is vanishing since $y_i$ converges to zero for all $i\in [n]$. Therefore, for all $n+1 \geq j \geq n+m$, $y_j$ also converges to zero.
 
 Thus, the origin of the model~\eqref{eq::sys_aux_y} is asymptotically stable. According to the 
 comparison principle, we conclude that the origin of the  model~\eqref{eq:errormatrix} is asymptotically stable for any non-zero initial condition. This finishes the proof.

  \subsection{Proof of Proposition \ref{prop:crate}}
 \label{app:crate}
 %The proof is simple. 
The Jacobian matrix at the origin of \eqref{eq::sys_sis_dt_z} is
$J(\mathbf{0},\mathbf{0})=\left[ 
        \begin{matrix}
	I+hB-hD & hB_{w}  \\
	hC_{w} & I-hD_w  \\
	\end{matrix}
        \right]=I+hB_f-hD_f.$
This indicates that, under the hypothesis of the proposition, the healthy state is locally exponentially stable with a convergence rate given by $s_1(I-hD_f+hB_f)$.

\subsection{Proof of Proposition \ref{prop:homo}}
 \label{app:homo}
 Equation \eqref{eq:end1} is now identical for all $i\in[n]$, while \eqref{eq:end2} is now identical for all $j\in[m]$.
Thus, we have $x^*_i=x^*$ and $w^*_j=w^*$.
$w^*=n\hat{c}x^*$ can be directly obtained by equation \eqref{eq:end2}.
Then, we plug it into equation \eqref{eq:end1} leading to the result. 
 
 \subsection{Proof of Proposition \ref{prop:pertubation}}
 \label{app:8}

We have that the endemic equilibrium is the solution of $\left( -hD_f+h(I-Z^*)B_f  \right)z^*=\mathbf{0}$ when $\rho(D_f^{-1}B_f)>1$.

Let us define a map $K:\mathbb R^{n+m}\times\mathbb  R^{(n+m)\times(n+m)}\times\mathbb  R^{(n+m)\times(n+m)}\to\mathbb R^{(n+m)}$ by $K(z^*,D_f,B_f)$ \\ $=\left( -hD_f+h(I-Z^*)B_f  \right)z^*$. Now, we show that the equation $K(z^*,D_f,B_f)=\mathbf{0}$ defines an function $z^*=g(D_f,B_f)$. Now, we consider a small perturbation $\Delta D_f$ and $\Delta B_f$ of $D_f$ and $B_f$ such that $\rho((D_f+\Delta D_f)^{-1}(B_f+\Delta B_f))>1$, where \begin{equation*}
    \begin{split}
\Delta B_{f}:=\left[ 
        \begin{matrix}
	\Delta B & \Delta B_{w}  \\
	\Delta C_{w} & \mathbf{0}  \\
	\end{matrix}
        \right],\\
        \Delta D_{f}:=\left[ 
        \begin{matrix}
	\Delta D & \mathbf{0}  \\
	\mathbf{0} & \Delta D_{w}  \\
	\end{matrix}
        \right].\\
    \end{split}
\end{equation*}

Let $z^*+\Delta z^*$ denote the endemic state resulting from the perturbation. Then, let $M=-h(D_f+\Delta D_f)$ and $P=h(I-(Z^*+\Delta Z^*))(B_f+\Delta B_f)$. We have 
\begin{equation} \label{eq::per1}
    \begin{split}
     &\left( M+P  \right)
     (z^*+\Delta z^*)
     =\mathbf{0}   \\
     &\left( -h(D_f+\Delta D_f)+h(I-Z^*)(B_f+\Delta B_f)  \right)z^*-h\Delta Z^*B_f z^*\\
     &- h\Delta Z^* \Delta B_f z^* 
     +\left( -h(D_f+\Delta D_f)\right.\\
     &\left.+h(I-(Z^*+\Delta Z^*))(B_f+\Delta B_f)  \right)(\Delta z^*)=\mathbf{0}
    \end{split}
\end{equation}

By using $K(z^*,D_f,B_f)=\mathbf{0}$ and ignoring the higher order of $\Delta$ term, equation \eqref{eq::per1} can be simplified as 
\begin{equation} \label{eq::per2}
\begin{split}
&\left( M+P \right)(z^*+\Delta z^*)   \\
     &=\left( -hD_f+hB_f-hZ^*B_f-B^* \right)\Delta z^*=\mathbf{0}   
\end{split}
\end{equation}
where $B^*=\Dg(B_f [x^*, \mathbf{0}]^\top)$.

We perform linearization around $z^*$,
\begin{equation} \label{eq::linearization}
\begin{split}
&\left( M+P  \right)(z^*+\Delta z^*)   \\
     =& \Dg(z^*) \Delta d_f +(Z^*-I) \Delta B_f z^*,
\end{split}
\end{equation}
where $\Delta d_f$ is the vector whose entry is the diagonal entry of $\Delta D_f$ and notice that $\Dg(z^*)$ is different from 
$Z^*=\left[ 
        \begin{matrix}
	\Dg(x^*) & \mathbf{0}  \\
	\mathbf{0} & \mathbf{0}  \\
	\end{matrix} \right]
	$. Thus, we have that
\begin{equation*}
    \left( -hD_f+hB_f-hZ^*B_f-B^* \right) z^*=-B^* z^*.
\end{equation*}
Since $B_f$ is a irreducible nonnegative matrix and  $z^* \gg \mathbf{0}$, $B^*$ is a positive diagonal matrix. Let $C>0$ such that $C<\min_{i,j} (B_f)_{ij}$. Then, $B^*>C I$ and 
$-B^*z^*<-Cz^*$. Notice that $-hD_f+hB_f-hZ^*B_f-B^*$ is an irreducible Metzler matrix, then $s_1(-hD_f+hB_f-hZ^*B_f-B^*)<-C<0$. Thus, $-hD_f+hB_f-hZ^*B_f-B^*$ is invertible and non-sigular. 
By the Implicit Function Theorem \cite{chiang1984fundamental} (pages 204-206), we have 
\begin{equation}
\begin{split}
    \Delta z^* &= (-hD_f+hB_f-hZ^*B_f-B^*)^{-1}\Dg(z^*) \Delta d_f\\ &+(-hD_f+hB_f-hZ^*B_f-B^*)^{-1}(Z^*-I) \Delta B_f z^*
\end{split}
\end{equation}

We also have the following lemma,
\begin{lemma} [Theorem 2.7 in Chapter 6 of \cite{berman1994nonnegative}]\label{lemma:hur}
Suppose that $M$ is a nonsingular, irreducible Hurwitz Metzler matrix.
Then, $M^{-1}\ll \mathbf{0}$.
\end{lemma}

By using Lemma \ref{lemma:hur}, we complete the proof.

\subsection{Proof of Theorem \ref{thm:6}}
\label{app:9}

 The proof is similar to the Theorem \ref{thm:hss}. 
 
 Firstly, we consider the case $s_{1}(I-hD_{f\min} + hB_{f\max}) < 1$. Let $M=I-hD_{f\min}+hB_{f\max}$
 By Lemma \ref{lem:greater1}, there exists a $P_1$ such that $M^\top P_1 M-P_1$ is negative definite. We define the Lyapunov function $V_1(z(t)) = (z(t))^\top P_1 (z(t))$. Let $\Bar{M}=I+h((I-Z(t))B_{f\max}-D_{f\min})$ and 
 $\hat{M}=I+h((I-Z(t))B_{f}(t)-D_{f}(t))$. 
 Thus, we have $\Bar{M}z(t)\geq \hat{M}z(t)$.
 
 \begin{equation}\label{eq::lyapunov2}
 \begin{split}
     &\Delta V_1(z(t+1))\\
     &=V_1(z(t+1))-V_1(z(t))\\
     &=  (z(t))^\top \hat{M}^{\top} P_1  \hat{M} (z(t))-(z(t))^\top P_1 (z(t))   \\
     &\leq (z(t))^\top \Bar{M}^{\top} P_1  \Bar{M} (z(t))-(z(t))^\top P_1 (z(t))\\
     &=(z(t))^\top ({M}^{\top} P_1  {M}-P_1) (z(t))\\
     &- 2h(z(t))^\top {B_f}^{\top} Z(t) P_1  {M} (z(t))\\
     &+h^2(z(t))^\top {B_f}^{\top} Z(t) P_1  Z(t) B_f (z(t))\\
     &\leq -2h(z(t))^\top {B_f}^{\top} Z(t) P_1  {M} (z(t))\\
     &+h^2(z(t))^\top {B_f}^{\top} Z(t) P_1  Z(t) B_f (z(t))
 \end{split}
 \end{equation}
 
The rest remains the same to the proof of theorem \ref{thm:hss}.

Then, we consider the case $s_{1}(I-hD_{f\min} + hB_{f\max}) = 1$. By Lemma \ref{lem:equal1}, there exists a $P_2$ such that $M^\top P_2 M-P_2$ is negative semi-definite. We define the Lyapunov function $V_2(z(t)) = (z(t))^\top P_2 (z(t))$. Then, we have 
\begin{equation*}
\begin{split}
    \Delta V_2(z(t+1)) &=  (z(t))^\top \hat{M}^{\top} P_2  \hat{M} (z(t))-(z(t))^\top P_1 (z(t))  \\
     &\leq (z(t))^\top \Bar{M}^{\top} P_2  \Bar{M} (z(t))-(z(t))^\top P_1 (z(t))\\
     &\leq -2h(z(t))^\top {B_f}^{\top} Z(t) P_2  {M} (z(t))\\
     &+h^2(z(t))^\top {B_f}^{\top} Z(t) P_2  Z(t) B_f (z(t))\\
     &\leq -h(z(t))^\top {B_f}^{\top} Z(t) P_2  {M} (z(t)) \\ &\leq 0.
\end{split}
\end{equation*}

The rest remains the same to the proof of theorem \ref{thm:hss}.

\subsection{Proof of Theorem \ref{thm:7}}
\label{app:10}

Since all pairs $(B_f(t),D_f(t))\in \Omega_{z^*}$ refer to the same endemic equilibrium, the error dynamics given by \eqref{eq::sys_sis_dt_xe} and \eqref{eq::sys_sis_dt_we} are still valid but with the time-varying parameters. They read as now

\begin{align}
     e_i(t+1) &= (1+\frac{h\delta_{i}(t)}{x_{i}^*-1})e_i(t)\notag \\
     &+(1-x_{i}^*-e_{i}(t))h(\sum_{j=1}^{n}\beta_{ij}(t)e_{j}(t)\\
     &+\sum_{j=1}^{m}\beta_{ij}(t)f_{j}(t),\label{eq::sys_sis_dt_xet} \\
     f_j(t+1)&= (1-h\delta_{j}^{w}(t))f_{j}(t)+h\sum_{k=1}^{n}c_{jk}^{w}(t)e_{k}(t).\label{eq::sys_sis_dt_wet}
\end{align}

It is equivalent to the error dynamics in matrix form

\begin{equation} \label{eq:errormatrix1}
    \hat{e}(t+1)=\phi(t)\hat{e}(t),
\end{equation}
where \begin{equation}
\hat{e}(t)=[e_{1}(t),\dots,e_{n}(t),f_1(t),\dots,f_m(t)]^{\top} 
\end{equation}
and
\begin{small}  
\begin{align} 
    &\phi(t)= \notag \\
    &\left[ 
        \begin{matrix}
	I-\Dg(\frac{h\delta_i(t)}{1-x_i^*})+\Dg(1-x(t))hB(t) & \Dg(1-x(t))hB_w (t) \\
	hC_w(t) & I-hD_{w} (t) \\
	\end{matrix}\right]. \label{eq:phi1}
\end{align}
\end{small}

Define \begin{equation} \label{eq:f1}
    F(t)=\left[ 
        \begin{matrix}
	I-diag(\frac{h\delta_i(t)}{1-x_i^*})+hB(t) & hB_w (t) \\
	hC_w(t) & I-hD_{w} (t) \\
	\end{matrix}\right], 
\end{equation}
the matrix $F$ is non-negative irreducible since all elements in $\Omega_{z^*}$ satisfy Assumption 2-6.

Let $\mu = [1,\frac{x^*_2}{x^*_1},\ldots,\frac{x^*_n}{x^*_1},\frac{w^*_1}{x^*_1},\dots, \frac{w^*_m}{x^*_1}]^{\top}$. Notice that
 \begin{equation}
  F(t) \mu = \mu.
 \end{equation}

 Since $F(t) $ is an irreducible nonnegative matrix and $\mu \gg \mathbf{0}$, it follows that $\rho(F(t)) = 1$ by Lemma \ref{lemma::nonneg_compare}. According to the Lemma \ref{lemma::perron_fr}, there exists a positive left eigenvector $v^{\top}$ such that $v^{\top} F (t) = v^{\top}$. Moreover, $B_f(t)$ is symmetric, which shows $F (t)$ is also symmetric. Thus, $v=\mu$.
 
 The following proof is similar to the proof of theorem \ref{thm:end}. Following the same proof, we get the conclusion under these two cases. 
 
  \emph{Case 1:} If $z(0) = \mathbf{0}$. We can directly obtain that the system will stay at the healthy state and doesn't converge to the endemic equilibrium. Moreover, $z(t)=0, \forall t \geq 0$ if and only if $z(0)=0$. 

\emph{Case 2:} If $z(0) \neq \mathbf{0}$, we show that there must exists a time $s$ such that 
 $x(s) \gg \mathbf{0}$. 
 
 Construct the following auxiliary system for any $k \geq s$. 
 \begin{equation} \label{eq::sys_aux_y1}
  y(t+1) = \Phi(t) y(t),
 \end{equation}
 with initial condition $y_i(s) = |e_i(s)|, \forall i \in \mathcal{V}$. We also have $\hat{e}(t+1)= \phi(t)\hat{e}(t)$. Since $\Phi(t)$ is nonnegative, it follows 
 that $-y(t) \leq \hat{e}(t)\leq y(t), \forall k \in \mathbb{N}$. 
 
 Define the Lyapunov function $V(t) = v^{\top}y(t)$, where $v$ is the positive left eigenvector of $F(t)$, which equals to $\mu$ since $F(t)$ is symmetric. We can calculate increment of the Lyapunov function as follows.
 \begin{equation}
  \begin{aligned}
   \Delta V(t) &= V(t+1)-V(t) = v^{\top} (\Phi(t)-I) y \\
               &= v^{\top}(\Phi(t)-F)y \\
               &= -h v^{\top} \left[ 
        \begin{matrix}
	\Dg(x(t)) & \Dg(x(t))  \\
	\mathbf{0} & \mathbf{0}  \\
	\end{matrix}\right]B_{f}(t) y \\
               &\leq 0. 
  \end{aligned}
 \end{equation}
 The remaining is the same to the proof of Theorem \ref{thm:end}.

 \subsection{Proof of Lemma \ref{lemma:boundmul}}
 \label{app:11}

We recall that the time-invariant system is given, element-wise, as follows
\begin{align}
     x_i^k(t+1)&= x_i^k(t)+h\left[ \left(1-\sum^{l}_{a=1}x^a_i(t)\right) \left(\sum_{j=1}^{n} \beta^k_{ij} x^k_j(t)\right.\right. \notag \\
 &\left.\left.+\sum_{j=1}^{m} \beta_{ij}^{wk} w^k_j(t)\right) -\delta_i^k x^k_i(t) \right], \label{eq::sys_sis_dt_xmul} \\
  w_j^k(t+1)&= w_j^k(t)+h\left(-\delta_{j}^{wk}w^k_{j}(t)+\sum_{a=1}^{n}c_{ja}^{wk}x^k_{a}(t)\right).\label{eq::sys_sis_dt_wmul}
\end{align} 
where all variables and parameters refer to the same meaning as the single-virus model and the subscript $k$ denotes that it is the variable or parameter of virus $k$.

%$x^k_{i}$ denotes the agent $i$'s infection probability of the $k$-th virus or the infection proportion of group $i$ to the $k$-th virus; $\beta^k_{ij} := \beta^k_{i} A^k_{ij}$, where $\beta_{i}^k$ is the infection rate of node $i$ to the $k$-th virus and $A_{ij}^k$ is the entry of the subgraph $\mathcal{G}^k$; $w_{j}^k$ represents the pathogen's concentration of the $k$-th virus in the $j$-th resource; $\delta_{j}^{wk}$ is decay rate of the $k$-th pathogen; $c_{jk}^{wk}$ is the effective $k$-th virus's person-resource contact rate per person or agent $j$ to the resource $k$.

We can rewrite equations \eqref{eq::sys_sis_dt_xmul} and \eqref{eq::sys_sis_dt_wmul} into the matrix form:

{\small\begin{equation} \label{eq::sys_sis_dt_xm1}
\begin{split}
     x^k(t+1)&= x^k(t)+h\left[\left(I-\sum_{a=1}^{l}\Dg(x^a(t))B^k-D^k\right)x^k(t)\right.\\
     &\left.+\left(I-\sum_{a=1}^{l}\Dg(x^a(t))\right)B^k_{w}w^k(t) \right],
\end{split}
\end{equation}} 
and
\begin{equation} \label{eq::sys_sis_dt_wm1}
 w^k(t+1)= w^k(t)+h\left(  -D^k_{w}w^k(t)+C^k_{w}x^k(t)   \right);
\end{equation}
where $x^k(t)=[x^k_{1}(t), \dots, x^k_{n}(t)]^{\top}$, $w^k(t)=[w^k_{1}(t), \dots, w^k_{m}(t)]^{\top}$, $B^k=[\beta^k_{ij}]_{n \times n}$, $B^k_w=[\beta^w_{ij}]_{n \times m}$, $D^k=\Dg(\delta^k_{1},\dots, \delta^k_{n})$, $D_{w}^k=\Dg(\delta_{1}^{wk},\dots, \delta_{n}^{wk})$ and $C_{w}^k=[c^{wk}_{jk}]_{m \times n}$.

We can further rewrite the systems~\eqref{eq::sys_sis_dt_xm1} and~\eqref{eq::sys_sis_dt_wm1} as follows:
\begin{equation} \label{eq::sys_sis_dt_zmul}
 z^k(t+1)= z^k(t)+h\left( -D^k_{f}+\left(I-\sum_{a=1}^{l}Z^a(t)\right)B^k_{f}\right)z^k(t),
\end{equation} 
where the matrix or vector of $z^k(t),Z^k(t),B^k_{f},D^k_{f}$ has the same structure and meaning of \eqref{eq::notation} and the subscript $k$ denotes that it is the matrix or vector of virus $k$.

Now, we can start our proof.

By our assumptions, all the parameters as well as the state variable $z(t)$ are all non-negative. So for $t=0$, we have $z^k(1)= z^k(0)+h\left( -D^k_{f}+(I-\sum_{a=1}^{l}Z^a(0))B^k_{f}\right)z^k(0) \\ \leq z^k(0)+h\left( -D^k_{f}+(I-Z^k(0))B^k_{f}\right)z^k(0)=\hat{z}^k(1)$. Next, suppose $z^k(t) \leq \hat{z}^k(t)$ holds. By an induction principle it follows that $z^k(t+1)\leq \hat{z}^k(t+1)$.

\subsection{Proof of Lemma \ref{lemma:inimul}}
\label{app:12}

$x_{i}^k(t) \in [0,1]$ and $w^k_{j}(t) \in [0,w^k_{\max}]$ is the direct consequence of lemma \ref{lemma:boundmul}. Then, we add all $x_{i}^k$ up and get
\begin{equation*} \label{eq::sys_sis_dt_xmulsum}
\begin{split}
 \sum_{k=1}^l x_i^k(t+1)&= \sum_{k=1}^l x_i^k(t)+h[ (1-\sum^{l}_{a=1}x^a_i(t)) \\
 &(\sum_{k=1}^l\sum_{j=1}^{n} \beta^k_{ij} x^k_j(t)+\sum_{k=1}^l\sum_{j=1}^{m} \beta_{ij}^{wk} w^k_j(t))\\ &-\sum_{k=1}^l\delta_i^k x^k_i(t) ]\\
 &\leq \sum_{k=1}^l x_i^k(t)+h[ (1-\sum^{l}_{a=1}x^a_i(t)) (\sum_{k=1}^l\sum_{j=1}^{n} \beta^k_{ij} \\
 &+\sum_{k=1}^l\sum_{j=1}^{m} \beta_{ij}^{wk} w^k_{\max} )
 -\min_{k\in[l]}(\delta_i^k)\sum_{k=1}^l x_i^k(t)] \\
 &=D(t).
\end{split}
\end{equation*}

Since $D(t)$ is the convex combination of $h\sum_{k=1}^{l}(\sum_{j=1}^{N} \beta^k_{ij} +\sum_{j=1}^{N} \beta_{ij}^{wk} w^k_{\max})\in [0,1]$ and $1-h\min_{k\in[l]}(\delta_i^k)\in [0,1)$, it follows that $D(t)\in [0,1]$.

\subsection{Proof of Lemma \ref{lemma:inimul1}}
\label{app:13}

It suffices to show that 
\begin{equation} \label{eq::sys_sis_dt_zmultleq1}
\begin{split}
    &z^k(t+1)= z^k(t)\\
    &+h\left( -D^k_{f}(t)+(I-\sum_{a=1}^{l}Z^a(t))B^k_{f}(t)\right)z^k(t)\\
    &\leq z^k(t)\\
    &+h\left( -D^k_{f\min}(t)+(I-\sum_{a=1}^{l}Z^a(t))B^k_{f\max}(t)\right)z^k(t)
\end{split}
\end{equation} 

The remaining is the same to the proof of Lemma \ref{lemma:inimul}.

\subsection{Proof of Lemma \ref{lemma:boundmul1}}
\label{app:14}

By our assumptions, all the parameters as well as the state variable $z(t)$ are all non-negative. So we have $z^k(1)= z^k(0)+h\left( -D^k_{f}(t)+(I-\sum_{a=1}^{l}z^a(0))B^k_{f}(t)\right)z^k(0) \\ \leq \hat{z}^k(1)=z^k(0)+h\left( -D^k_{f}(t)+(I-Z^k(0))B^k_{f}(t)\right)z^k(0)$. Suppose $z^k(t) \leq \hat{z}^k(t)$ holds, it follows that $z^k(t+1)\leq \hat{z}^k(t+1)$. It completes the proof.

\subsection{Proof of Theorem \ref{thm:hmul}}
\label{app:15}
By Lemma \ref{lemma:boundmul}, $z(t)$ is upper bounded by its single-virus counterpart. Since $s_1(I-D_f^k + B_f^k) \leq 1$, its single-virus counterpart converges to zero by Theorem \ref{thm:hss}. Moreover, $z(t)$ is also lower bounded by $0$. According to the comparison principle, all $z^k(t)$ converges to zero for any arbitrary initial condition.

\subsection{Proof of Theorem \ref{thm:mul1}}
\label{app:16}
For any $a\neq k$, following the same proof as in Theorem \ref{thm:hmul}, $z^a(t)$ converges to zero. If $z^k(0)=\mathbf{0}$, then the $k$-th virus is eradicated and the system converges to the healthy state.

If $z^k(0)\neq \mathbf{0}$, we have
\begin{align*}
 z^k(t+1)&= z^k(t)+h\left( -D^k_{f}+(I-Z^k(t))B^k_{f}\right)z^k(t) \\
 &-\sum_{a\neq k}^{l}Z^a(t)B^k_{f}z^k(t)
\end{align*} 
where $-\sum_{a=1}^{l}Z^a(t)B^k_{f}z^k(t)$ can be regarded as a vanishing perturbation, which converges to zero when $t \rightarrow \infty$. Thus, $z^k(t)$ can be seen as an autonomous system at the end. Then  From theorem \ref{thm:end}, the autonomous system converges to the unique endemic state $z^{k*} \gg \mathbf{0}$. Thus, it completes the proof.

\subsection{Proof of Theorem \ref{thm:coexist}}
\label{app:17}
Firstly, we need the following Lemma.

\begin{lemma} \label{lemma:cdeqb}
Consider a continuous-time model $\Dot{x}=f(x)$ and its discrete-time counterpart using Euler's Method $x(t+1)=x(t)+hf(x(t))$. Both models share the same equlibra.
\end{lemma}

\begin{pf}
It suffices to notice that $\Dot{x}=f(x^*)=0$ shares the same solution with $x^*=x^*+ h f(x^*)$. 
\end{pf}

\begin{lemma} \label{lemma:cdeqb1}
Consider a continuous-time model $\Dot{x}=f(x)$ and its discrete-time counterpart using Euler's Method $x(t+1)=x(t)+hf(x(t))$. Let $x^*$ be the equilibrium of both systems. Let the Jacobian matrix of system $\Dot{x}=f(x)$ at $x^*$ be $J^1(x^*)$.
\begin{itemize}
    \item [i)]$x^*$ is locally unstable for the discrete-time model if $x^*$ is locally unstable for the continuous-time model.
    \item [ii)]$x^*$ is locally exponentially stable for the discrete-time model if $h< \frac{2}{\rho(J^1(x^*))}$ and $x^*$ is locally exponentially stable for the continuous-time model.
\end{itemize}
 
\end{lemma}

\begin{pf}
 Let the Jacobian matrix of the corresponding discretized system $x(t+1)=x(t)+hf(x(t))$ at $x^*$ be denoted by $J^2(x^*)$. Thus, $J^2(x^*)=hJ^1(x^*)+I$, which implies that the eigenvalues of $J^2(x^*)$ follow $\lambda(J^2(x^*))=h\lambda(J^1(x^*))+1$, where by $\lambda(\cdot)$ we denote the spectrum. We can see that an eigenvalue $\lambda(J^1(x^*))$ located on the right-side of the complex plane corresponds to an eigenvalue $\lambda(J^2(x^*))$ outside the unit circle. Thus, $x^*$ is locally unstable for the discrete-time model if $x^*$ is locally unstable for the continuous-time model. On the other hand, if $h< \frac{2}{\rho(J^1(x^*))}$, then an eigenvalue $\lambda(J^1(x^*)$ on the left-side of the complex plane corresponds to an eigenvalue $\lambda(J^2(x^*))$ within the unit circle.
\end{pf}

Consider the continuous-time model \cite{janson2020networked} when $l=2$ and $m=1$. We compare the conditions with that in \cite[Theorem 5]{janson2020networked} . If Assumptions \ref{ass:multi}-\ref{ass:h1m}, then Assumption 1 in \cite{janson2020networked} hold and $s_1(I-hD_f^k + hB_f^k) > 1$ is equivalent to $s_1(hD_f^k - hB_f^k) > 0$ in Theorem 5 of \cite{janson2020networked}. Thus, by lemma \ref{lemma:cdeqb} and \cite[Theorem 5]{janson2020networked}, the result follows.

\subsection{Proof of Theorem \ref{thm:win}}
\label{app:18}
The existence of these 3 equilibra can be derived by lemma \ref{lemma:cdeqb} of this paper and theorem 6, lemma 8 of \cite{janson2020networked}.

Now, we prove the local stability for each equilibrium. 

Notice that $J$ defined in \eqref{eq:J} is the Jacobian matrix of the continuous-time counterpart proposed in \cite{janson2020networked}.
By directly applying the lemma \ref{lemma:cdeqb1}, we achieve theorem \ref{thm:win}.

\subsection{Proof of Theorem \ref{thm:mul11}}
\label{app:19}
By Lemma \ref{lemma:boundmul1}, $z(t)$ is upper bounded by its single-virus counterpart. Since $s_1(I-D_{f\min}^k + B_{f\max}^k
) \leq 1$, its single-virus counterpart converges to zero by Theorem \ref{thm:hss}. Moreover, $z(t)$ is also lower bounded by 0. According to the comparison principal, all $z^k(t)$ converges to zero for any arbitrary initial condition.

\subsection{Proof of Theorem \ref{thm:mul12}}
\label{app:20}
For any $a\neq k$, following the same proof as in theorem \ref{thm:mul11}, $z^a(t)$ converges to zero. If $z^k(0)=\mathbf{0}$, then the $k$-th virus is eradicated and the system converges to the healthy state.

If $z^k(0)\neq \mathbf{0}$, $z^k$ can be seen as an autonomous system
\begin{equation}
 z^k(t+1)= z^k(t)+h\left( -D^k(t)_{f}+(I-Z^k(t))B^k(t)_{f}\right)z^k(t)
\end{equation} 
with a vanishing perturbation $-\sum_{a=1}^{l}Z^a(t)B^k_{f}(t)z^k(t)+Z^k(t)B^k_{f}(t)z^k(t)$, which converges to zero when $t \rightarrow \infty.$ From theorem \ref{thm:end}, the autonomous system converges to the unique endemic state $z^{k*} \gg \mathbf{0}$. Thus, it completes the proof.

\end{document}